\documentclass[usenatbib]{mnras}
\usepackage[T1]{fontenc}
\DeclareRobustCommand{\VAN}[3]{#2}
\let\VANthebibliography\thebibliography
\def\thebibliography{\DeclareRobustCommand{\VAN}[3]{##3}\VANthebibliography}
\usepackage{graphicx}
\usepackage{amsmath}
\usepackage{amssymb}
\usepackage{color}
\usepackage{url}
\usepackage{dsfont}
\usepackage{subfig}
\usepackage{pdflscape}
\usepackage{changepage}
\usepackage{CJKutf8}

\usepackage{threeparttable}

\newcommand\lsim{\mathrel{\rlap{\lower4pt\hbox{\hskip1pt$\sim$}}
        \raise1pt\hbox{$<$}}}
\newcommand\gsim{\mathrel{\rlap{\lower4pt\hbox{\hskip1pt$\sim$}}
        \raise1pt\hbox{$>$}}}
\newcommand{\lya}{\ifmmode\mathrm{Ly}\alpha\else{}Ly$\alpha$\fi}
\newcommand{\lyb}{\ifmmode\mathrm{Ly}\beta\else{}Ly$\beta$\fi}
\newcommand{\igm}{\ifmmode\mathrm{IGM}\else{}IGM\fi}
\newcommand{\lae}{\ifmmode\mathrm{LAE}\else{}LAE\fi}
\newcommand{\h}{\ifmmode\mathrm{H}\else{}H\fi}
\newcommand{\hi}{\ifmmode\mathrm{H\,{\scriptscriptstyle I}}\else{}H\,{\scriptsize I}\fi}
\newcommand{\hii}{\ifmmode\mathrm{H\,{\scriptscriptstyle II}}\else{}H\,{\scriptsize II}\fi}
\newcommand{\cmb}{\ifmmode\mathrm{CMB}\else{}CMB\fi}
\newcommand{\qso}{\ifmmode\mathrm{QSO}\else{}QSO\fi}
\newcommand{\eor}{\ifmmode\mathrm{EoR}\else{}EoR\fi}
\newcommand{\heii}{\ifmmode\mathrm{He\,{\scriptscriptstyle II}}\else{}He\,{\scriptsize II}\fi}
\newcommand{\heiii}{\ifmmode\mathrm{He\,{\scriptscriptstyle III}}\else{}He\,{\scriptsize III}\fi}
\newcommand{\ciii}{\ifmmode\mathrm{C\,{\scriptscriptstyle III]}}\else{}C\,{\scriptsize III]}\fi}
\newcommand{\oiii}{\ifmmode\mathrm{O\,{\scriptscriptstyle III}}\else{}O\,{\scriptsize III}\fi}
\newcommand{\aliii}{\ifmmode\mathrm{Al\,{\scriptscriptstyle III}}\else{}Al\,{\scriptsize III}\fi}
\newcommand{\mgii}{\ifmmode\mathrm{Mg\,{\scriptscriptstyle II}}\else{}Mg\,{\scriptsize II}\fi}
\newcommand{\fe}{\ifmmode\mathrm{Fe}\else{}Fe\fi}
\newcommand{\nv}{\ifmmode\mathrm{N\,{\scriptscriptstyle V}}\else{}N\,{\scriptsize V}\fi}
\newcommand{\niv}{\ifmmode\mathrm{N\,{\scriptscriptstyle IV]}}\else{}N\,{\scriptsize IV]}\fi}
\newcommand{\cii}{\ifmmode\mathrm{C\,{\scriptscriptstyle II}}\else{}C\,{\scriptsize II}\fi}
\newcommand{\civ}{\ifmmode\mathrm{C\,{\scriptscriptstyle IV}}\else{}C\,{\scriptsize IV}\fi}
\newcommand{\siv}{\ifmmode\mathrm{Si\,{\scriptscriptstyle IV}}\else{}Si\,{\scriptsize IV}\fi}
\newcommand{\siii}{\ifmmode\mathrm{Si\,{\scriptscriptstyle II}}\else{}Si\,{\scriptsize II}\fi}
\newcommand{\siiii}{\ifmmode\mathrm{Si\,{\scriptscriptstyle III]}}\else{}Si\,{\scriptsize III]}\fi}
\newcommand{\ovi}{\ifmmode\mathrm{O\,{\scriptscriptstyle VI}}\else{}O\,{\scriptsize VI}\fi}
\newcommand{\sioiv}{\ifmmode\mathrm{Si\,{\scriptscriptstyle IV}\,\plus O\,{\scriptscriptstyle IV]}}\else{}Si\,{\scriptsize IV}\,+O\,{\scriptsize IV]}\fi}

\newcommand{\avenf}{$\bar{x}_{\rm HI}$}

\newcommand{\intermediateHII}{\textsc{\small Intermediate \hii{}}}

\newcommand{\cmfst}{\textsc{\small 21CMFAST}}

\usepackage{newtxtext,newtxmath}

\pdfoutput=1

\makeatletter
\providecommand\phantomcaption{\caption@refstepcounter\@captype}
\makeatother

\title[IGM damping wing constraints with XQR-30]{IGM damping wing constraints on the tail end of reionisation from the enlarged XQR-30 sample}
\author[B. Greig et al.] {Bradley~Greig$^{1,2,3}$\thanks{E-mail:~greigb@unimelb.edu.au}, A. Mesinger$^{4}$, E. Ba{\~n}ados$^{5}$, G. D. Becker$^{6}$, S. E. I. Bosman$^{7,5}$, \newauthor 
H. Chen$^{8}$, F. B. Davies$^5$, V. D’Odorico$^{9,4,10}$, A.-C. Eilers$^{11}$, S. Gallerani$^{4}$, \newauthor 
M. G. Haehnelt$^{12,13}$, L. Keating$^{14}$, S. Lai$^{15}$, Y. Qin$^{1,3}$, E. Ryan-Weber$^{16,3}$, S. Satyavolu$^{17}$, \newauthor 
F. Wang$^{18}$, J. Yang$^{18}$ \& Y. Zhu$^{18,7}$\\
$^1$School of Physics, University of Melbourne, Parkville, VIC 3010, Australia \\
$^2$Research School of Astronomy \& Astrophysics, Australian National University, Canberra, ACT 2611, Australia \\
$^3$ARC Centre of Excellence for All-Sky Astrophysics in 3 Dimensions (ASTRO 3D) \\
$^4$Scuola Normale Superiore, Piazza dei Cavalieri, I-56126 Pisa, Italy \\
$^5$Max-Planck-Institut f{\"u}r Astronomie, K{\"o}nigstuhl 17, D-69117 Heidelberg, Germany \\
$^6$Department of Physics and Astronomy, University of California, Riverside, CA 92521, USA \\
$^7$Institute for Theoretical Physics, Heidelberg University, Philosophenweg 12, D-69120, Heidelberg, Germany \\
$^8$Canadian Institute for Theoretical Astrophysics, University of Toronto,60 St George St, Toronto, ON M5R 2M8, Canada\\
$^9$INAF–Osservatorio Astronomico di Trieste, Via G.B. Tiepolo, 11, I-34143, Trieste, Italy \\
$^{10}$IFPU–Institute for Fundamental Physics of the Universe, via Beirut 2, I34151 Trieste, Italy \\
$^{11}$MIT Kavli Institute for Astrophysics and Space Research, Massachusetts Institute of Technology, Cambridge, MA 02139, USA \\
$^{12}$Institute of Astronomy, University of Cambridge, Madingley Road, Cambridge CB3 0HA, UK \\
$^{13}$Kavli Institute of Cosmology, University of Cambridge, Madingley Road, Cambridge CB3 0HA, UK \\
$^{14}$Institute for Astronomy, University of Edinburgh, Blackford Hill, Edinburgh, EH9 3HJ, UK \\
$^{15}$Commonwealth Scientific and Industrial Research Organisation (CSIRO), Space \& Astronomy, P. O. Box 1130, Bentley, WA 6102, Australia \\
$^{16}$Centre for Astrophysics and Supercomputing, Swinburne University of Technology, Hawthorn, Victoria 3122, Australia \\
$^{17}$Tata Institute of Fundamental Research, Homi Bhabha Road, Mumbai 400005, India \\
$^{18}$Steward Observatory, University of Arizona, 933 N Cherry Ave, Tucson, AZ 85721, USA
}

\pubyear{2024}

\begin{document}
\label{firstpage}
\pagerange{\pageref{firstpage}--\pageref{lastpage}}
\begin{CJK}{UTF8}{gkai} 
\maketitle
\end{CJK}
\begin{abstract}
\noindent
The attenuation of Ly$\alpha$ photons by neutral hydrogen in the intergalactic medium (IGM) at $z\gtrsim5$ continues to be a powerful probe for studying the epoch of reionisation. Given a framework to estimate the intrinsic (true) Ly$\alpha$ emission of high-$z$ sources, one can infer the ionisation state of the IGM during reionisation. In this work, we use the enlarged XQR-30 sample of 42 high-resolution and high-SNR QSO spectra between $5.8\lesssim\,z\lesssim\,6.6$ obtained with VLT/X-Shooter to place constraints on the IGM neutral fraction. This is achieved using our existing Bayesian QSO reconstruction framework which accounts for uncertainties such as the: (i) posterior distribution of predicted intrinsic \lya{} emission profiles (obtained via covariance matrix reconstruction of the Ly$\alpha$ and \nv{} emission lines from unattenuated high-ionisation emission line profiles; \civ{}, \sioiv{} and \ciii{}) and (ii) distribution of ionised regions within the IGM using synthetic damping wing profiles drawn from a $1.6^3$ Gpc$^3$ reionisation simulation. Following careful quality control, we used 23 of the 42 available QSOs to obtain constraints/limits on the IGM neutral fraction during the tail-end of reionisation. Our median and 68th percentile constraints on the IGM neutral fraction are: $0.20\substack{+0.14\\-0.12}$ and $0.29\substack{+0.14\\-0.13}$ at $z = 6.15$~and 6.35. Further, we also report 68th percentile upper-limits of $\bar{x}_{\hi{}} < 0.21$, 0.20, 0.21 and 0.18 at $z = 5.8, 5.95, 6.05$~and 6.55. These results imply reionisation is still ongoing at $5.8\lesssim\,z\lesssim\,6.55$, consistent with previous results from XQR-30 (dark fraction and Ly$\alpha$ forest) along with other observational probes considered in the literature.
\end{abstract} 
\begin{keywords}
cosmology: theory -- observations -- dark ages, reionisation, first stars -- early Universe -- galaxies: intergalactic medium -- quasars: emission lines
\end{keywords}

\section{Introduction}

The epoch of reionisation (EoR) denotes the final major baryonic phase change of the Universe, when the pervasive cold and dense neutral hydrogen fog in the intergalactic medium (IGM) is evaporated by the intense glow of UV ionising radiation from the first sources. The morphological evolution of the IGM proceeds in a patchy manner whereby individual ionised (\hii{}) regions form first around the earliest structures before growing and merging with their nearest neighbours, percolating and eventually ionising almost all of the neutral hydrogen in the IGM. Observing and subsequently characterising the EoR is of fundamental importance as it will advance our knowledge of the nature of the first structures to form in the Universe.

Our most promising observational tool for probing the EoR is the detection of the 21-cm hyperfine spin-flip transition of neutral hydrogen via radio interferometry. However, this continues to remain elusive with the best-yet upper-limits on the 21-cm signal \citep{Mertens:2020,Trott:2020,Abdurashidova:2022b} still a few orders of magnitude away from our theoretical expectations \citep{Ghara:2020,Greig:2020,Mondal:2020,Ghara:2021,Greig:2021,Abdurashidova:2022,HERA:2023}.

In the absence of directly measuring the neutral hydrogen in the IGM, we can instead infer its presence via its impact on the \lya{} photons emitted by bright astrophysical sources such as galaxies and QSOs. At lower redshifts, one can directly infer the ionisation state of the IGM by measuring the absorption of \lya{} photons within the \lya{} forest along the line-of-sight from the source. However, at $z\gtrsim5$ the scattering cross-section of \lya{} photons is sufficiently large that even trace amounts of neutral hydrogen ($x_{\hi} \gtrsim 10^{-4}$) are enough to produce completely saturated transmission \citep{Fan:2006p4005}. As a result, fluctuations in the neutral fraction owing to the patchy nature of reionisation are almost indistinguishable from fluctuations post-reionisation of the ultraviolet background, density or temperature \citep[e.g.][]{D'Aloisio:2015,Keating:2018}. However, physical models predict different large-scale fluctuations and biases of these fields, which can be used to constrain the EoR history and galaxy properties (see for example figure 2 in \citealt{Qin:2021}).

Alternatively, a more robust probe of the IGM neutral fraction is to measure the imprint of the \lya{} damping wing (e.g. \citealt{Rybicki1979,MiraldaEscude:1998p1041}). This approach takes advantage of the Lorentzian wings away from the resonant central core of the \lya{} absorption cross-section. In these wings, the amplitude of the scattering cross-section is several orders of magnitude lower and imprints a smooth absorption profile away from the central saturated absorption. Thus, even a completely neutral IGM will impart a smooth, measurable imprint in the transmission spectrum of the background source sufficiently redward from the line centre.

This IGM damping wing signature has been successfully measured to obtain constraints on the IGM neutral fraction using both galaxies and QSOs. For QSOs, their intrinsic brightness allows the damping wing imprint to be inferred from individual spectra \citep[e.g.][]{Mesinger:2007,Mortlock:2011p1049,Bolton:2011p1063,Greig:2017,Banados:2018,Davies:2018,Wang:2020,Yang:2020}. On the other hand, for galaxies, the extraction of the damping wing signal requires averaging over a sufficiently large statistical sample owing to the fainter nature of galaxies \citep[e.g.][]{Mesinger:2015p1584,Mason:2018,Hoag:2019,Mason:2019,Umeda:2023}, however, it can be more difficult to distinguish the IGM component from the host galaxy environment than is the case for QSOs \citep[e.g.][]{Heintz:2023,Keating:2023,Keating:2023b}.

Importantly, the key requirement for extracting the IGM damping wing imprint is a robust method to predict the intrinsic \lya{} emission, given we inherently observe an attenuated spectrum. For QSOs, numerous methods exist in the literature which differ considerably in their methodology \citep[e.g.][]{Davies:2018a,Dominika:2020,Fathivavsari:2020,Reiman:2020,Bosman:2021,Chen:2022,Liu:2021,Sun:2022} but fundamentally boil down to predicting the \lya{} profile from unattenuated spectral information redward of \lya{}. In \citet{Greig:2024}, a detailed comparison of the various reconstruction pipelines in the literature, focussing on the reconstruction of the \lya{} line profile, are performed as part of a blind reconstruction challenge. This extends on the work of \citet{Bosman:2021}, comparing methods to predict the placement of the \lya{} forest continuum.

In this work, we use the Bayesian reconstruction pipeline initially introduced in \citet{Greig:2017a} which predicts the intrinsic \lya{} profile by drawing from an N-dimensional normal distribution with a covariance matrix containing the correlations between the \lya{} and \nv{} line and the unattenuated high ionisation emission lines (\civ{}, \sioiv{} and \ciii{}). This approach assumes that the emission lines can be modelled as either a single or double component Gaussian profile characterised entirely by the width, height and velocity offset. We then couple this reconstruction framework to distributions of synthetic IGM damping wings drawn from a large-volume EoR simulation \citep{Mesinger:2016p6167}. In doing so, we account for the statistical uncertainties that arise since we are using single lines of sight to each QSO spectra to place constraints on the volume averaged IGM neutral fraction during the EoR.

Previously, IGM damping wing analyses have focussed solely on the highest redshift QSOs to be discovered at $z\gtrsim7$ \citep{Mortlock:2011p1049,Banados:2018,Wang:2020,Yang:2020}\footnote{Although, see \citet{Dominika:2024} who very recently performed an analysis of a much larger sample of 18 QSOs from $6.0 < z < 7.1$}. However, in this work we take advantage of the Ultimate X-shooter legacy survey of QSOs at $5.8 \lesssim z \lesssim 6.6$, XQR-30 \citep{DOdorico:2023}. This programme obtained 30 high signal-to-noise (S/N) QSO spectra at high resolution ($\sim30$ km s$^{-1}$). These were supplemented by a further 12 high quality $z\sim6$ QSOs from the X-shooter archive to produce a final sample of 42 QSOs. The sheer size of this QSO sample and the corresponding redshift coverage will enable unique constraints on the tail-end of the EoR to be recovered.

This work is structure as follows. In Section~\ref{sec:data} we provide a brief description of the observational data and in Section~\ref{sec:Method} we describe our full analysis pipeline. In Section~\ref{sec:Results} we provide the main results along with detailed discussions. In Section~\ref{sec:Conclusion} we conclude with our closing remarks. Unless stated otherwise, we quote all quantities in co-moving units and adopt the cosmological parameters:  ($\Omega_\Lambda$, $\Omega_{\rm M}$, $\Omega_b$, $n$, $\sigma_8$, $H_0$) = (0.69, 0.31, 0.048, 0.97, 0.81, 68 km s$^{-1}$ Mpc$^{-1}$), consistent with recent results from the Planck mission \citep{PlanckCollaboration:2016p7780}.

\section{Data} \label{sec:data}

\subsection{Sample}

The data used in this work come from the XQR-30 sample \citep{DOdorico:2023}, which is an ESO Large Programme ``XQR-30: the Ultimate XSHOOTER legacy survey of quasars at $z \simeq 5.8 - 6.6$" (P.I. V. D'Odorico; 1103.A-0817). Targeting 30 of the brightest known QSOs within this redshift range with X-shooter \citep{Vernet:2011}, we obtained a median resolution of $R\sim9500-13700$ in the visible arm and $R\sim7600-11000$ in the near-infrared arms. For this work, we re-binned the observed QSO spectra to 50 km/s. In addition to these 30 QSOs, a further 12 QSOs with similar redshift range, magnitude, resolution and S/N to the original XQR-30 sample were added from the X-shooter archive to yield the enlarged XQR-30 sample. This entire sample was reduced using a custom pipeline \cite[][]{Lopez:2016,Becker:2019}, with minor modifications made for XQR-30 \citep[see][for further details]{DOdorico:2023}.

\subsection{Final damping wing sample} \label{sec:sample}

Unfortunately not all of the 42 QSOs from the enlarged XQR-30 sample can be used for IGM damping wing studies. For example, we must avoid QSOs which exhibit absorption by dense, neutral absorption systems in close proximity to the host QSO (for example, Damped \lya{} systems; DLAs) which produce their own damping wing imprint which can imitate the damping wing signal from a partially neutral IGM \citep[e.g.][]{Prochaska:2008,Banados:2019,Davies:2023}. Further, QSOs which exhibit broad absorption lines (BALs) must also be excluded as these lines either contaminate the observed flux around \lya{} and \nv{} or impact our ability to accurately measure the continuum and emission line properties of the QSOs redward of \lya{} which are pre-requisite for our intrinsic \lya{} reconstruction pipelines. In Table~\ref{tab:XQR-30} we summarise the full XQR-30 sample, highlighting which QSOs are retained for our damping wing studies, and discuss the exclusions of certain objects in greater detail below.

\begin{table*}
\caption{A summary of the enlarged XQR-30 dataset: Columns correspond to the QSO name, redshift, emission line used for obtaining the redshift, recovered IGM neutral fraction (68 percentiles, either represented as an upper-limit or constraint) and brief comments highlighting the reason for each QSOs exclusion or retention (see main text for further details).}
\begin{tabular}{@{}lcccc}
\hline
QSO name & redshift ($z$) & redshift method & $\bar{x}_{\hi{}}$ & Comments   \\
\hline
PSO J029-29 & 5.976 & \mgii{} & $<0.22$ & \\
VST-ATLAS J029-36 & 6.013 & \mgii{} & $<0.40$ & \\
VDES J0224-4711 & 6.525 & \mgii{} & $<0.33$ & \\
PSO J060+24 & 6.170 & \mgii{} & $0.29\substack{+0.24 \\ -0.19}$ & \\
VDES J0408-5632 & 6.033 & \mgii{} & $0.24\substack{+0.19 \\ -0.15}$ & mini BAL or weak BAL$^a$ \\
PSO J108+08 & 5.955 & \mgii{} & $<0.31$ & \\
SDSS J0842+1218 & 6.0754 & [\cii{}] & $<0.30$ & mini BAL or weak BAL$^a$ \\
PSO J158-14 & 6.0685 & [\cii{}] & $<0.32$ & \\
PSO J217-16 & 6.1498 & [\cii{}] & $0.29\substack{+0.20 \\ -0.18}$ & \\
PSO J231-20 & 6.5869 & [\cii{}] & $<0.29$ & mini BAL or weak BAL$^a$ \\
PSO J242-12 & 5.840 & \mgii{} & $<0.25$ & \\
PSO J308-27 & 5.799 & \mgii{} & $0.33\substack{+0.22 \\ -0.20}$ & \\
PSO J323+12 & 6.5872 & [\cii{}] & $0.23\substack{+0.19 \\ -0.15}$ & \\
PSO J359-06 & 6.1722 & [\cii{}] & $<0.34$ & \\
SDSS J0100+2802 & 6.3269 & [\cii{}] & $0.41\substack{+0.19 \\ -0.18}$ & \\
VST-ATLAS J025-33 & 6.3373 & [\cii{}] & $<0.33$ & \\
ULAS J0148+0600 & 5.977 & \mgii{} & $0.32\substack{+0.23 \\ -0.20}$ & mini BAL or weak BAL$^b$ \\
PSO J036+03 & 6.5405 & [\cii{}] & $<0.31$ & \\
SDSS J0836+0054 & 5.773 & \mgii{} & $<0.33$ & \\
SDSS J0927+2001 & 5.7722 & [\cii{}] & $<0.33$  & \\
SDSS J1030+0524 & 6.304 & \mgii{} & $<0.40$  & \\
SDSS J1306+0356 & 6.0330 & [\cii{}] & $<0.36$ & \\
CFHQS J1509-1749 & 6.1225 & [\cii{}] & $<0.38$ & \\      
\hline
\vspace{0.8mm}
PSO J007+04 & 6.0015 & [\cii{}] & - & pDLA \\
PSO J009-10 & 6.0040 & [\cii{}] & - & BAL \\
PSO J023-02 & 5.817 & \mgii{} & - & BAL, pDLA \\
PSO J025-11 & 5.816 & \mgii{} & - & pDLA \\
PSO J065-26 & 6.1871 & [\cii{}] & - & pDLA \\
PSO J065+01 & 5.804 & \mgii{} & - & mini BAL or weak BAL \\
PSO J089-15 & 5.972 & \mgii{} & - & BAL \\
J0923+0402 & 6.6330 & [\cii{}] & - & BAL \\
PSO J183+05 & 6.4386 & [\cii{}] & - & pDLA \\
PSO J183-12 & 5.893 & \mgii{} & - & mini BAL or weak BAL\\
PSO J217-07 & 6.166 & \mgii{} & - & BAL \\
DELS J1535+1943 & 6.358 & \mgii{} & - & Heavily reddened \\
PSO J239-07 & 6.1102 & [\cii{}] & - & BAL, pDLA \\
VDES J2211-3206 & 6.3394 & [\cii{}] & - & BAL \\ 
VDES J2250-5015 & 5.985 & \mgii{} & - & BAL \\
SDSS J2310+1855 & 6.0031 & [\cii{}] & - & mini BAL or weak BAL, pDLA \\
WISEA J0439+1634 & 6.5188 & [\cii{}] & - & BAL \\
SDSS J0818+1722 & 5.960 & \mgii{} & - & pDLA \\
ULAS J1319+0950 & 6.1347 & [\cii{}] & - & pDLA \\
\hline
\end{tabular}
\footnotesize{\\$^a$ Although identified as a possible mini/weak BAL by \citet{Bischetti:2023}, the associated BAL complex does not impact the \lya{}+\nv{} emission line complex. $^b$ a potential weak BAL complex however the large blueshift should not contaminate the \lya{} + \nv{} emission line complex \citep{Becker:2015}.}
\label{tab:XQR-30}
\end{table*}

\citet{Bischetti:2022} performed an extensive \civ{} BAL analysis of the XQR-30 sample, with a follow up analysis exploring \nv{}, \sioiv{} and \mgii{} BAL systems \citep{Bischetti:2023}. Following this, 14 BAL QSOs were identified. However, of these, three (VDES J0408-5632, SDSS J0842+1218 and PSO J231-20) were deemed suitable for our damping wing analysis as the velocities of the BAL systems do not contaminate the \lya{} + \nv{} line complex. In D'Odorico et al. (in prep.) proximate DLA systems (pDLAs) have been identified within 7 of the XQR-30 QSOs, however only 4 of these are unique (i.e are not also identified as BAL QSOs). We further exclude the heavily reddened QSO J1535+1943, which is thought to be obscured \citep{Yang:2021}.

Of the additional 12 archival X-shooter spectra in the XQR-30 sample, two exhibit proximate DLA systems (J0818+1722 and J1319+0950). Further, J0439+1634 is known to exhibit a BAL system. J0148+0600 also likely exhibits a BAL system \citep{Becker:2015} however it is highly blue shifted and should not contaminate the \lya{} + \nv{} line complex.

In total, this results in 23 QSOs from the XQR-30 sample that we deem suitable for extracting the IGM damping wing imprint.

\section{Method} \label{sec:Method}

Our damping wing analysis relies on the covariance matrix reconstruction pipeline introduced by \citet{Greig:2017a} and subsequently extended in \citet{Greig:2022} in order to simultaneously reconstruct the \lya{} and \nv{} emission lines. This is combined with a Bayesian sampling of synthetic IGM damping wing profiles drawn from large-volume EoR simulations to be able to infer the IGM neutral fraction. Below, we summarise the key components of our full analysis pipeline and refer the reader to the aforementioned works for more in-depth discussions.

\subsection{Intrinsic Ly$\alpha$ profile reconstruction} \label{sec:Reconstruction}

In \citet{Greig:2017a} we introduced a method to reconstruct the intrinsic \lya{} line profile of QSOs using a covariance matrix of correlations between \lya{} and other measurable high ionisation lines (namely, \civ{}, \sioiv{} and \ciii{}). This covariance matrix was determined from a carefully selected (visually inspected) training set of 1673 moderate-$z$ ($2.08 < z < 2.5$), high signal to noise (S/N $>15$) QSOs from SDSS-III (BOSS) DR12 \citep{Dawson:2013p5160,Alam:2015p5162} to avoid the presence of BAL and DLA systems. In \citet{Greig:2022} we expanded the emission line covariance matrix to allow the simultaneous reconstruction of \lya{} and \nv{}.

The underlying assumption of this pipeline is that each emission line can be modelled by a sum of Gaussian profiles, each fully described by its amplitude, width and velocity offset away from systemic. In \citet{Greig:2017a} we performed basic model selection to determine that the \lya{} and \civ{} line profiles are best characterised by a two component Gaussian, consisting of both a broad and narrow component. In addition to the Gaussian emission lines, we model the QSO continuum as a single two-parameter power-law ($\propto\lambda^{\alpha_{\lambda}}$) over our entire fitting region (1180-2300\AA) and normalise all QSO spectra to unity at 1450\AA. For all remaining emission lines, a single component Gaussian profile was preferred\footnote{Note, we fit all observed emission lines that fall within this specific wavelength region, but only use the high-ionisation lines for the covariance matrix. That is, we also simultaneously fit for the following lines \siii{}, O\,{\scriptsize I}/\siii{}, \cii{}, \heii{}, \oiii{}/Al\,{\scriptsize II}/Fe\,{\scriptsize II} and \aliii{} (see \citealt{Greig:2017a} for further details). Further, we do not consider an emission component from an Fe continuum, however, for the purposes of our studies we find such a component unlikely to impact on our QSO fits.}. This results in a $21\times21$ covariance matrix of emission line parameters that is used to construct a 21-dimensional Normal distribution which describes the full properties of our 5 emission lines. The two parameters describing the power-law continuum do not enter the covariance matrix as they do not correlate with the parameters describing the emission lines.

\subsubsection{Enlarged QSO training set}

For this work we significantly expand our original training set of QSOs, updating our training set to use the BOSS DR16Q database \citep{Lyke:2020}. We both expand the redshift range from our original $2.08 < z < 2.5$ to a broader $2.08 < z < 4.0$ and reduce our S/N cut down from S/N $>15$ to S/N $>6.5$. Owing to this expansion of the training set, we now forego visual inspection of the QSOs in our training set. The primary reason for expanding this training set is two fold: (i) to avoid biasing our training set based on the quality of QSO spectra (large amplitude \lya{} peaks etc.) and (ii) to increase the sampled distribution of emission line parameters to ensure a larger region of validity for our covariance matrix.

After broadening our training set, we are left with $\sim55,000$ QSOs. We then reduce this sample further by removing QSOs with too many coincident absorption features\footnote{Absorption features are identified through an automated procedure based on a rolling 2\AA\ window, flagging all flux bins below $3\sigma$ ($5\sigma$ around \lya{}) and then roughly determining the absorber line centre which is assigned a Gaussian profile (see \citealt{Greig:2017a} for further details). All absorption features are simultaneously fit to the QSO spectrum along with the emission lines, varying their width, amplitude and velocity offset.} around our emission lines which can make determining the true profile problematic. Thus, we restrict our sample to QSOs with fewer than 20 absorption features across our fitting range of 1180-2300\AA\ and fewer than 8 in the 1180-1260\AA\ region around \lya{} and \nv{}. While these choices are arbitrary, they strike a balance between providing a sufficient number of QSOs for evaluating our covariance matrix and limiting the number of outlier spectra. Note, in fitting these absorption features we do not distinguish by their origin (i.e. \lya{} forest, intervening metal absorption and/or QSO associated absorption). Simply, from experience we find that for an increasing number of individual absorption lines within the 1180-1260\AA\ region, determining the intrinsic \lya{} profile from the training QSOs becomes problematic as the \lya{} components become strongly degenerate with these numerous absorption features. Beyond just the \lya{} region, we additionally find that too many absorption lines can impact our ability to fit the two component power-law continuum, impacting all subsequent emission line fits. After these cuts, our final training set consists of 30,166 QSOs, corresponding over an order of magnitude increase from the 1673 QSOs used in \citet{Greig:2017a}. In Appendix~\ref{sec:covariance} we provide the correlation coefficient matrix demonstrating the various emission line correlation strengths. With such a large number of QSOs in the training set, it is unlikely that our resultant correlations will be strongly biased. Interestingly, despite substantially increasing the size of our QSO training set for this work, we find very similar correlations strengths between our emission lines as we have found previously \citep{Greig:2022}. Although the correlation strengths remain similar, the notably larger training set demonstrates that these line correlations exist over broader ranges for the individual emission line parameter properties.

\subsubsection{Reconstruction pipeline for XQR-30} \label{sec:rec_pipe}

With our updated covariance matrix of emission line parameters, we now outline our method for reconstructing the intrinsic \lya{} and \nv{} emission lines. The key steps are as follows:
\begin{itemize}
\item The observed rest-frame QSO spectrum is fit over $\lambda =1275-2300$\AA\ using the systemic QSO redshifts\footnote{Note, our BOSS QSO training set is converted to rest-frame using the provided pipeline redshift, rather than the [\cii{}] or \mgii{} redshifts of the XQR-30 sample. Although this may lead to biases in the reconstructions due to differences in the obtained red/blueshifts from systemic, this is likely a sub-dominant effect relative to the large scatter in the reconstructed \lya{} + \nv{} line and of the synthetic IGM damping wings \citep[see][for detailed discussions]{Greig:2019}} outlined in Table~\ref{tab:XQR-30}.
\item We jointly fit the two-parameter power-law continuum and the high-ionisation lines \sioiv{}, \civ{} (2-component) and \ciii{} emission line profiles using our Gaussian profiles. Additionally, to improve the overall fit to the observed spectrum, we simultaneously allow a variable number of absorption lines, each modelled as a single Gaussian profile, to be fit.
\item Using the corresponding fits to the \civ{}, \sioiv{} and \ciii{} emission lines (amplitude, width and velocity offset) of our observed high-$z$ QSO spectrum, we use the maximum $a$-posteriori values for these lines and evaluate our 21-dimensional Normally distributed covariance matrix model. This allows us to reduce the full 21-parameter model down to a 9 dimensional estimate of just the joint \lya{} and \nv{} intrinsic emission line profile (two Gaussian components for \lya{} and one for \nv{}).
\item We then draw joint intrinsic \lya{} and \nv{} profiles from this 9 dimensional covariance matrix. We further utilise the amplitude of the observed spectrum redward of 1250\AA\ as an additional prior, to ensure the reconstructed profiles match the observed flux within this range.
\end{itemize}

\subsection{Modelling the IGM damping wing during reionisation} \label{sec:Damping}

In order to extract the imprint of a partially neutral IGM, we compare the observed signal against our intrinsic profile reconstructions convolved with synthetic IGM damping wing profiles. Our synthetic IGM damping wing profiles are obtained from the Evolution of 21-cm Structure (EOS; \citealt{Mesinger:2016p6167})\footnote{http://homepage.sns.it/mesinger/EOS.html} 2016 simulations. These semi-numerical reionisation simulations have side lengths of 1.6~Gpc over 1024 voxels and include state-of-the-art sub-grid prescriptions for inhomogeneous recombinations and the photo-heating suppression of star-formation. Further, through judicial choices of star-formation efficiency and the typical masses of halos hosting star-forming galaxies, three distinct EoR morphologies are explored.

Following \citet{Greig:2022}, we only consider the \intermediateHII\ model, which corresponds to reionisation driven by galaxies residing in $M_h \gsim 10^9 M_\odot$ haloes\footnote{In \citet{Greig:2019} we explored the impact that different EoR morphologies had on the recovered IGM neutral fraction constraints, finding only weak evidence for morphology dependence. The median IGM neutral fraction due to the different EoR morphologies was found to be only $\pm0.05$ relative to the typical 68th percentile uncertainty of $\pm0.15-0.20$. This should not impact our results in any discernible way.  }. This model is consistent with recent results from forward-modelling the \lya{} forest opacity distributions and ultra-violet (UV) galaxy luminosity functions \citep[LFs;][]{Qin:2021}.

We statistically account for the fact we only obtain a single observed QSO sightline by drawing a large sample of synthetic IGM damping wing profiles varying both: (i) the sightline orientation (selecting 10 different lines of sight through the IGM centred on the host halo) and (ii) the diversity of host QSO environments in the IGM. In total, we draw $10^5$ synthetic IGM damping wing profiles obtained from the $10^4$ largest identified halos (corresponding to roughly $6\times10^{11} < M_h < 3\times10^{12}$ $M_\odot$)\footnote{Although the XQR-30 QSOs are known to be bright and powered by massive blackholes \citep{Mazzucchelli:2023}, they are consistent with the range of halo masses considered within our semi-numerical simulations and thus should not strongly impact on our inferred constraints on the IGM neutral fraction.} and 10 randomly oriented sightlines originating from the central host halo. To have these synthetic damping wing profiles varied by IGM neutral fraction we take the ionisation fields obtained from different redshift snapshots from the simulation, which assumes the halo locations do not significantly change due to bulk motions across the different redshift snapshots. Following this, we obtain 24 unique, unevenly spaced values of the \igm{} neutral fraction, corresponding to the original redshift sampling of our semi-numerical simulations (i.e. 24 different snapshots during the EoR), spanning from $\bar{x}_{\hi{}} = 0.02$~to 0.95 for our analysis.

These synthetic damping wing profiles are constructed by assuming a minimum local \hii{} bubble size, $R_{\rm min}$, which for the $z>7$ QSOs \citep{Greig:2017,Greig:2019,Greig:2022} was adopted to be their corresponding proximity zone size. This is necessary since within our EoR simulations we do not model the impact that the QSO has on growing its own local \hii{} bubble. The actual size of the local \hii{} region is the combination of the cumulative ionising output of the resident local galaxies (which is incorporated within our simulations) and the unknown QSO contribution, which is crudely implemented by assuming that at a minimum this QSO plus galaxy contribution has to have at least ionised the size of the measured QSO proximity zone. The contribution of the QSO to the local \hii{} region is strongly dependent on its accretion history, obscuration and emission geometry all of which are highly uncertain and thus difficult to impose physically motivated priors. Therefore, in this work we only search for the IGM damping wing imprint between $\lambda=1218-1230$\AA\ to specifically avoid having to simultaneously model the host QSO environment blueward of \lya{}, thus minimising additional modelling uncertainties. This choice makes our approach more agnostic to the QSO emission model and its associated priors. The lower bound of 1218\AA\ ensures we are sufficiently far from the gravitational potential of the halo, avoiding potential signatures of inflowing gas. This approach differs to that of \cite{Davies:2018} who utilise numerical simulations to develop a hybrid scheme allowing for the search of the damping wing imprint over an extended region of $\lambda\sim1190-1230$\AA. This enables these QSO emission model uncertainties such as the QSO lifetime and environment to be folded into their analysis. A detailed discussion of the consequences of these assumptions are provided in \citet{Greig:2022} but in summary our redward approach implicitly imposes a complex prior on the QSO lifetime, but should not strongly impact our results on the inferred neutral fraction. A quantitative estimate of the impact of this prior is beyond the scope of this work, as it requires a detailed comparison of the two different damping wing extraction methods.

Due to the significantly increased number of QSOs in the XQR-30 sample relative to our previous $z>7$ QSO studies and the additional computational burden of having to regenerate these synthetic sightlines for each possible $R_{\rm min}$, we instead choose to only use the existing synthetic damping wing profiles from the four previously analysed $z>7$ QSOs. Therefore, for each XQR-30 QSO we simply average over our existing data, drawing $10^5$ profiles for $R_{\rm min}$; $\sim11-16$ comoving Mpc\footnote{Note that this is slightly smaller than some of the proximity zone sizes measured for the XQR-30 sample by \citet{Satyavolu:2023}; 2-7 proper Mpc at $z=5.8-6.6$. However, unfortunately we no longer have the base simulation data to extend our upper limit of $R_{\rm min}$. Nevertheless, this should not significantly impact our results due to the broad range of modelling uncertainties.}. By averaging over these $R_{\rm min}$ we effectively broaden our resultant posteriors, leading to a more conservative estimate of the IGM neutral fraction. In Appendix~\ref{sec:prior} we explore the impact of assuming different $R_{\rm min}$ for one of our QSOs in our sample. Between our lower and upper limits on this minimum radii we find the inferred IGM neutral fraction shifts by $\sim0.02$, which is considerably smaller than the width of the resultant posterior. Therefore, this choice of convolving over $R_{\rm min}$ should have little impact on our inferred results on the IGM neutral fraction.

\subsection{Initial QSO reconstruction performance} \label{sec:init}

\begin{figure*} 
	\begin{center}
	  \includegraphics[trim = 0.9cm 0.8cm 0cm 0.6cm, scale = 0.4]{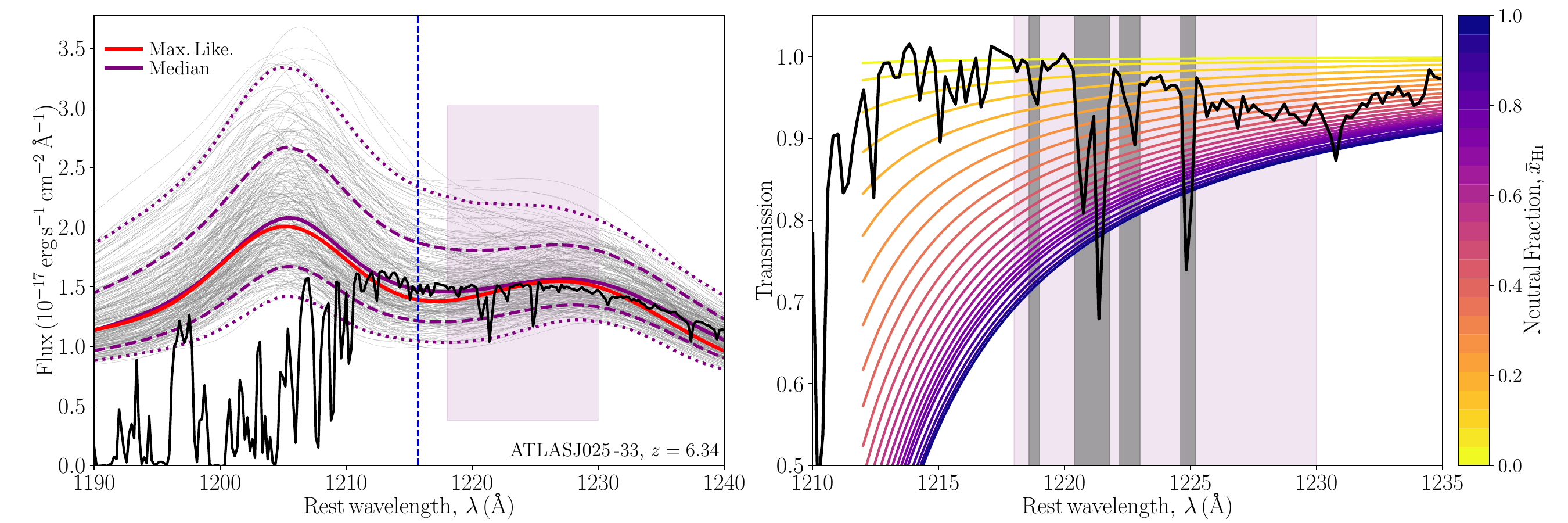}
	\end{center}
\caption[]{A visual demonstration of the initial covariance reconstruction for ATLAS J025-33 at $z=6.34$ highlighting problems with our original reconstruction pipeline. \textit{Left panel:} The observed QSO spectrum (black curve), corresponding maximum-likelihood (red curve) reconstruction and 300 random profile draws (grey curves) from our full reconstructed covariance matrix. The solid purple curve corresponds to the median reconstruction profile obtained from the full distribution of reconstructed profiles while the dashed (dotted) curves correspond to the 68th (95th) distribution of profiles. The vertical blue dashed line corresponds to rest-frame \lya{} while the purple box denotes the region over which we fit for the IGM damping wing imprint (1218-1230\AA). \textit{Right panel:} The observed \lya{} transmission profile (black curve) obtained by dividing the observed spectrum by the median reconstructed profile. The coloured lines correspond to the median synthetic IGM damping wing profiles for each IGM neutral fraction value sampled by our EoR simulations. Grey shaded regions correspond to absorption features that are masked by our damping wing fitting pipeline.}
\label{fig:uncalibrated}
\end{figure*}

Our initial attempts at performing intrinsic reconstructions following the procedure outlined in Section~\ref{sec:rec_pipe} for the XQR-30 sample proved to be problematic. As a visual demonstration, in the left panel of Figure~\ref{fig:uncalibrated} we provide our initial attempt at reconstructing the intrinsic profile of ATLAS J025-33. The black curve corresponds to the observed spectrum, while the red line corresponds to the maximum-likelihood profile reconstruction. The grey lines correspond to 300 random posterior draws of our reconstructed profiles, while the purple solid line corresponds to the median profile and the dashed (dotted) purple lines demarcate the 68th and 95th percentile ranges determined over the full distribution of reconstructed profiles. Finally, the purple shaded box corresponds to the interval over which we extract the smooth IGM damping wing component (1218-1230\AA).

In this example, the maximum likelihood reconstruction (also the median, but to a lesser extent as it is obtained over the distribution instead of an individual profile) drops below the observed spectrum between rest-frame 1210-1220\AA\footnote{The reconstructed profile also drops below the observed spectrum at $>1235$\AA, however, this is less prevalent across our sample and occurs outside of the region over which we fit for the damping wing (i.e. 1218-1230\AA). Nevertheless, this likely is related to the issues with the width and placement of the \nv{} line from our reconstructed profile.}. Owing to the strong blueshift of this particular QSO, this corresponds to the flux between the \lya{} and \nv{} emission lines. Importantly, this `feature' is not just restricted to one QSO, instead it is persistent across the full QSO sample, systematically occurring between the \lya{} and \nv{} emission lines. In the right panel of Figure~\ref{fig:uncalibrated} we demonstrate the impact that this behaviour can have on our recovery of the smooth IGM damping wing imprint. Here, we show the \lya{} transmission profile (black curve), obtained by dividing the observed spectrum (black curve, left panel) by the median reconstruction profile (purple curve, left panel). The coloured curves correspond to the median synthetic IGM damping wing profiles for each value of the IGM neutral fraction sampled by our dataset (increasing neutral fraction from top to bottom, see Section~\ref{sec:Damping}). The smooth, monotonically increasing synthetic damping wing profiles are incompatible with the decreasing amplitude of the observed \lya{} transmission within the 1218-1230\AA\ region owing to this underestimate of the intrinsic flux. Attempting to jointly fit the observed \lya{} transmission and synthetic damping wing profiles within this region was problematic for our pipeline.

\subsection{Calibrating our QSO reconstruction pipeline} \label{sec:recal}

We initially assumed that the origin of this feature was the higher quality of the X-Shooter sample relative to the training set of SDSS (BOSS) QSOs used in the construction of our covariance matrix. The higher resolution would then lead to more prominent emission line features (less blended), in particular around \lya{} and \nv{} driving this difference. Subsequently, we degraded the XQR-30 QSOs to better mimic the spectral properties of the QSOs within our training set but found that this behaviour of a reduction of flux between \lya{} and \nv{} persisted. 

Therefore, in order to establish the origin of this feature in our profile predictions we performed a reconstruction of QSOs drawn from two different samples both at $3.5 < z < 4.5$ which should be unaffected by IGM attenuation (i.e. where we can easily establish the true intrinsic profile shape). For this, we constructed two samples of 30 QSOs each, one observed with X-shooter to obtain QSOs of similar quality and properties of the XQR-30 sample and one with SDSS (BOSS DR16Q)\footnote{Although some QSOs from our expanded training set likely exist in this sample of 30 BOSS QSOs, since our reconstruction method relies on drawing from a Normal distribution based on a covariance matrix of line properties our reconstructions will not be biased by their appearance.} to be representative of the quality of our training set spectra. For the former, we utilise the XQ100 sample \citep{Lopez:2016}. Importantly, by considering two samples of QSOs from distinctly different instruments, we can determine whether: (i) the problems in our reconstruction are intrinsic to the quality of the spectrum and (ii) whether it is acceptable for a reconstruction method trained on lower quality spectra to be applied to any instrument.

Originally, the scope of this investigation was to be limited to just our covariance matrix reconstruction pipeline. However, it quickly morphed into a robust and detailed comparison of all available reconstruction pipelines in the literature. For detailed discussions resulting from this comparison we refer the reader to \citet{Greig:2024}. In short, the issue was identified to be intrinsic to our covariance reconstruction method (see below) and that the quality of the spectra in the training set (e.g. resolution or S/N) relative to the observed spectra did not impact the reconstruction performance. In the remainder of this section, we focus only on the results intrinsic to our covariance matrix reconstruction pipeline, relevant for this work.

\begin{figure*} 
	\begin{center}
	  \includegraphics[trim = 0.15cm 0.9cm 0cm 0.5cm, scale = 0.438]{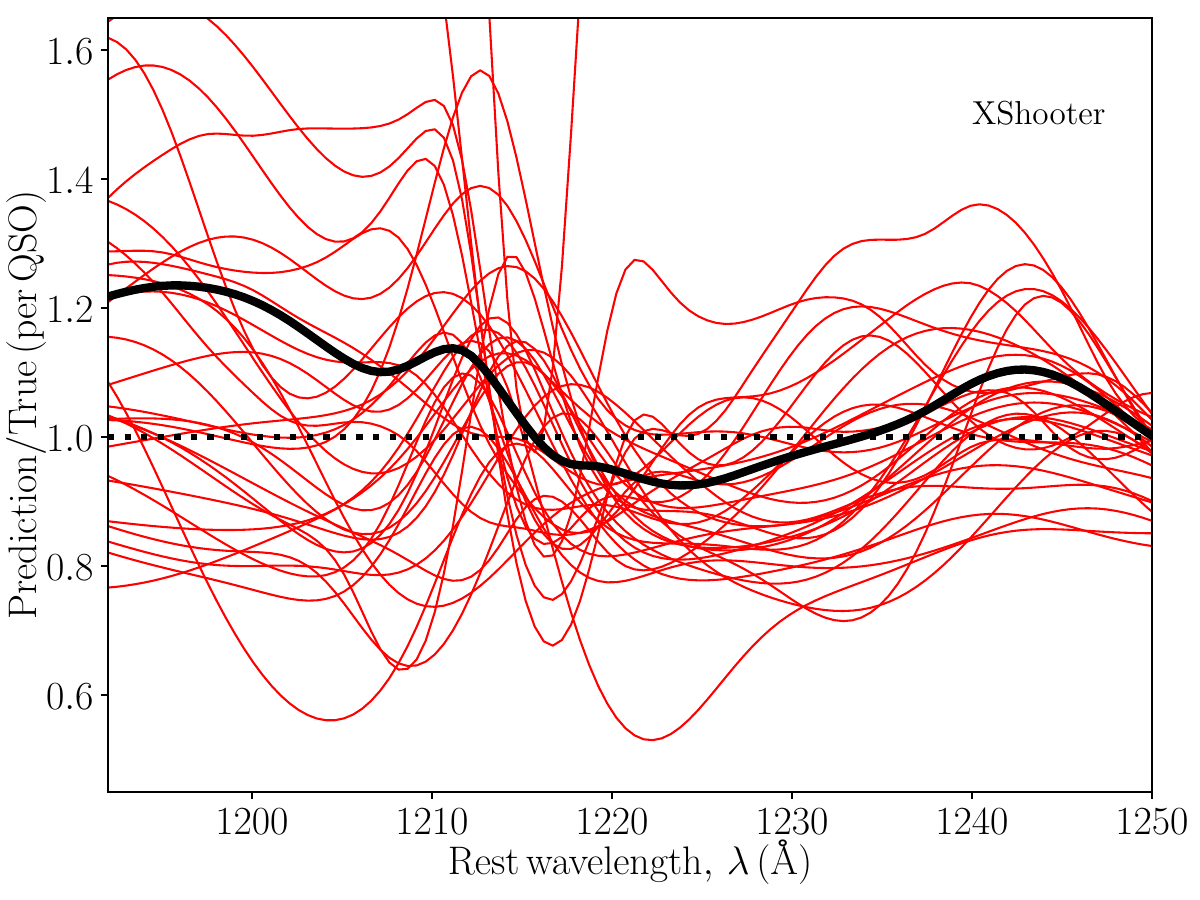}
	  \includegraphics[trim = 0.35cm 0.9cm 0cm 0.5cm, scale = 0.438]{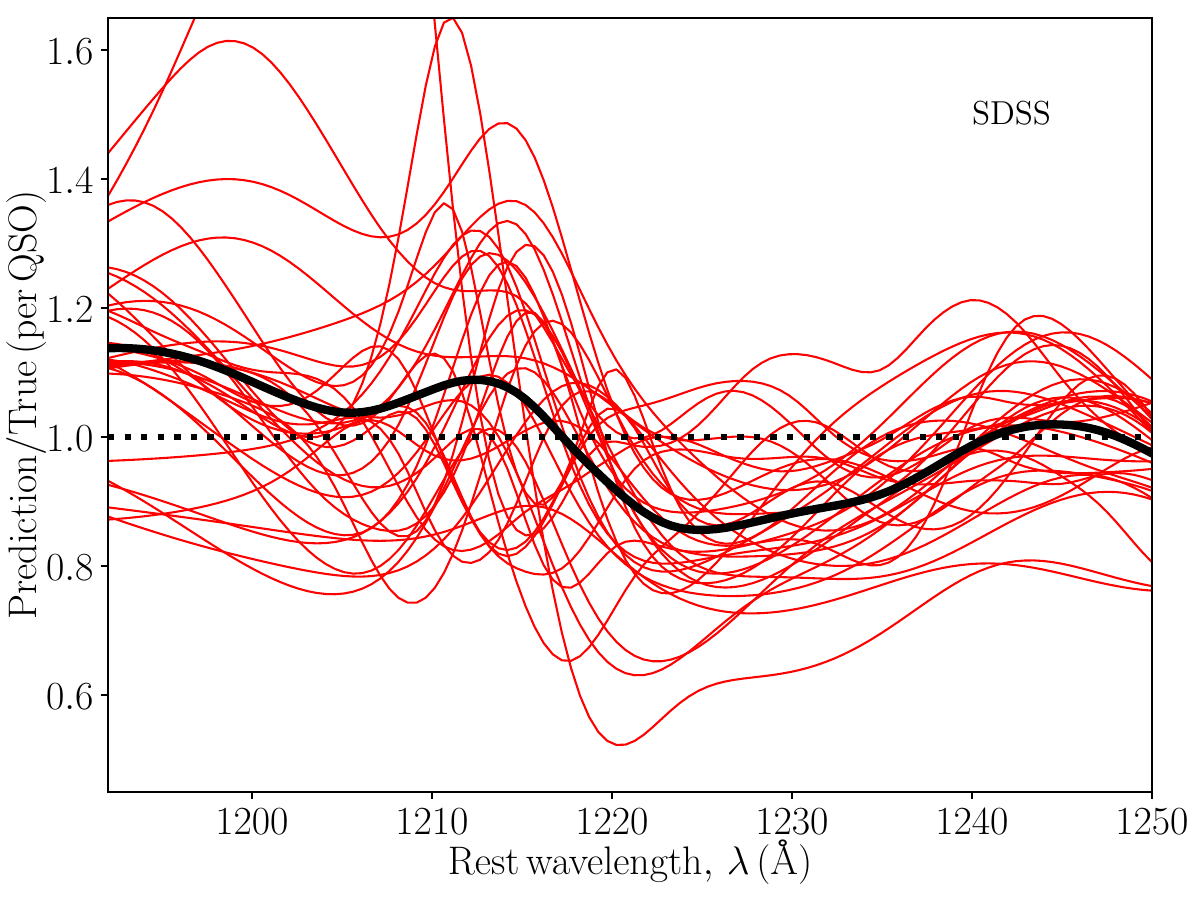}
	\end{center}
\caption[]{The performance of our covariance matrix reconstruction method from two distinct samples of 30 QSOs from X-shooter (left panel) and SDSS (BOSS; right panel) at $3.5 < z < 4.5$. The red solid lines correspond to the flux ratio (reconstructed maximum-likelihood profile over true intrinsic profile) for each individual QSO in the samples. The black solid lines correspond to the median determined over each sample.}
\label{fig:Correction}
\end{figure*}

In Figure~\ref{fig:Correction} we summarise the performance of our covariance matrix reconstruction pipeline across the two distinct samples of 30 QSOs. The left panel corresponds to the X-shooter sample, while the right panel corresponds to the SDSS (BOSS) sample. Here, the red curves correspond to the flux ratios (reconstructed maximum-likelihood profile divided by true, unattenuated flux) for each individual QSO in each sample. The black solid line corresponds to the median flux ratio obtained over the 30 QSOs in each sample. For both, we follow the reconstruction procedure outlined in Section~\ref{sec:rec_pipe}.

Immediately from Figure~\ref{fig:Correction} it is evident that the flux ratios across both samples follow very similar behaviour, exemplified by the remarkably similar shape in the median flux ratios, albeit with the SDSS sample exhibiting a slightly larger offset. This highlights that the higher quality X-shooter spectra in comparison to the BOSS training set does not produce any additional source of systematic biases. Thus explaining why the degradation of the XQR-30 spectra did not alleviate the issue. Further, for both samples a clear dip is evident between 1215 and 1230\AA\footnote{Note that this feature occurs over a different rest-frame wavelength range purely due to the blueshift of ATLAS J025-33 used in our previous example.}. That is, our reconstruction pipeline consistently underestimates the shape of the reconstructed \lya{} + \nv{} line profile within this region by $\sim10$~per cent irrespective of the quality of the observed spectra being reconstructed. Thus the problematic feature identified earlier is systematic to our reconstruction pipeline.

The origin of this behaviour is due to the modelling of the QSO flux within this specific wavelength region. Here, the QSO flux is the sum of 3 Gaussian components\footnote{Although in theory we could consider additional Gaussian components to improve the ability to fit the \lya{} + \nv{} line complex, the overall correlation strengths between additional (beyond two) \lya{} components and the other high ionisation lines would diminish as additional components are not always required to fit the observed training data.}, the sum of a broad and narrow component for \lya{} and a single component for \nv{}. Deviations in the predicted locations and profile shapes of these Gaussian components can produce this decrement in the QSO flux between \lya{} and \nv{}. For example, a larger predicted separation between the broad \lya{} and \nv{} lines caused by either (or both) a bluer than expected \lya{} line centre or redward \nv{} line centre or a narrower than expected width in these line profiles would produce a flux decrement within this spectral region. Owing to the notably lower strength of the \nv{} emission line correlations (compared to the \lya{} components) with the other emission lines in our covariance matrix (see e.g. Figure~\ref{fig:cov}) the \nv{} is going to be less strongly constrained (i.e. larger uncertainty) and thus more likely to be the cause of the flux decrement than the broad \lya{} component. Importantly, given that these components are sufficient for fitting QSOs within this spectral region (i.e. in our training set), it may also indicate that there are higher order correlations between our emission line parameters we are missing within our covariance matrix approach. In future, we aim to explore this further but for the purposes of this work, we will re-calibrate our profiles to mitigate this feature.

It is worth pointing out that this feature was not identified in our previous study of $z>7$ QSOs for several reasons. Other than for ULASJ1120+0641, there is no clear/prominent \nv{} line in the data and further a very broad \lya{} line is preferred. As a result, there was limited opportunity for this feature to manifest as it requires a notable \nv{} line component. For ULASJ1120+0641, the relatively broad observed \nv{} line profile extends sufficiently redward to benefit from our prior on the observed QSO flux. Therefore, this prior enables us to predict a relatively broad \nv{} line component, limiting the potential impact of this feature. Thus, only with the exquisite quality of the X-shooter spectra coupled with the more prevalent \nv{} line features has this feature become apparent.

It is important to note that this particular problem is intrinsic to our reconstruction method, as it relies upon the sum of three predicted Gaussian components to describe the \lya{} + \nv{} line complex. Other reconstruction pipelines in the literature provide flux predictions based off the sum of PCA components or profiles drawn from machine learning approaches which are more agnostic to the direct emission line correlations \citep[e.g.][]{Davies:2018a,Dominika:2020,Fathivavsari:2020,Reiman:2020,Bosman:2021,Chen:2022,Liu:2021,Sun:2022}. As a result, these appear to be less susceptible to such systematic biases in the reconstructed profiles (see \citealt{Greig:2024} for more detailed discussions).

Crucially, and relevant for this work, the consistency of the flux ratios across the two samples implies that we should be able to recalibrate our reconstruction pipeline to correct for this systematic offset in our reconstruction profiles. To do so, we convolve each random draw from our covariance matrix reconstruction by the distribution of flux ratios shown in Figure~\ref{fig:Correction} (both from the X-Shooter and SDSS sample). A consequence of this recalibration is that the distribution of reconstructed profiles will become broader, reflecting the increased scatter injected by sampling from these flux ratios and subsequently also broader uncertainties on the inferred IGM neutral fraction. Note, this convolution only occurs once we have corrected for the blueshift of the reconstructed QSO. That is, we shift these flux ratios to be centred on the location of the \lya{} profile estimated from the reconstruction pipeline to ensure we are correctly recalibrating over the problematic region between \lya{} and \nv{}.

\begin{figure*} 
	\begin{center}
	  \includegraphics[trim = 0.9cm 0.8cm 0cm 0.6cm, scale = 0.4]{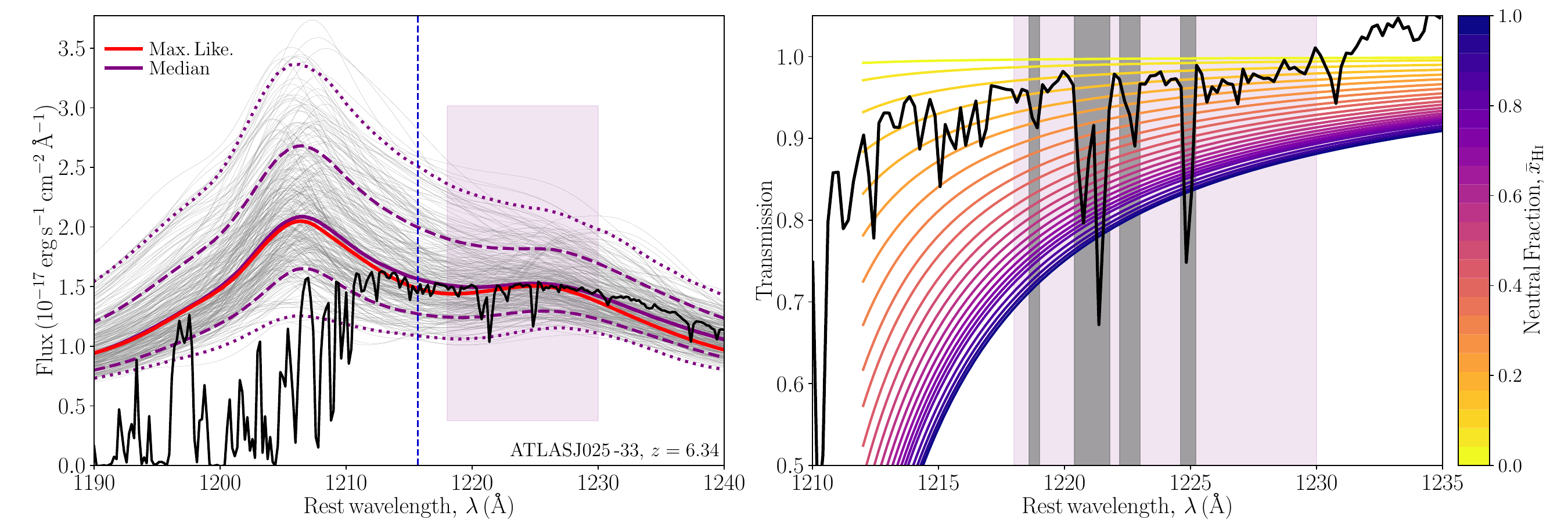}
	\end{center}
\caption[]{The same as Figure~\ref{fig:uncalibrated} except after recalibrating our reconstruction pipeline using template spectra from two different QSO samples (see text for further details).}
\label{fig:Calibration}
\end{figure*}

In Figure~\ref{fig:Calibration} we demonstrate the performance of this recalibration on the QSO, ATLAS J025-33, (as shown in Figure~\ref{fig:uncalibrated}). In the left panel, we present our calibrated reconstruction and in the right panel, the corresponding \lya{} transmission following this recalibration. Note, for the maximum-likelihood profile (red curve) we simply multiply it by the median flux ratio from Figure~\ref{fig:Correction} for visualisation purposes. In the left panel, we can easily identify the two main aspects of this recalibration procedure. Firstly, the shape of the median profile between 1210 and 1220\AA\ no longer possesses the parabolic dip observed in Figure~\ref{fig:uncalibrated} (more readily visible in the right panel) and secondly, the broader distribution (and increased diversity) of the random profiles draws from our covariance matrix reconstruction. Note, that by performing our recalibration for this particular object, the predicted median flux is now below the observed spectrum at $\gtrsim1230$\AA. However, this is not too concerning as it is at the redward edge of the \nv{} line, whereas the purpose of the recalibration is to improve the reconstructions more so on the blue side of \nv{} and the redward edge of \lya{} (the joint contribution of the \lya{} and \nv{} line components). Nevertheless, this will be explored in greater detail in future work when working to improve on this QSO reconstruction method.

The right panel of Figure~\ref{fig:Calibration} more readily demonstrates the improvements this recalibration brings to our reconstruction pipeline. Correcting this systematic offset in the shape of the reconstructed profiles produces corresponding \lya{} transmission profiles which now better match the expected shape of the synthetic IGM damping wing profiles. That is, the observed \lya{} transmission is now monotonically increasing with rest-frame wavelength, enabling more robust template matching against the synthetic sampling wing profiles drawn from the EoR simulations.

\subsection{Joint fitting to obtain IGM neutral fraction constraints} \label{sec:JointFitting}

We obtain our constraints on the IGM neutral fraction by jointly sampling our synthetic damping wing profiles from our EoR simulations convolved with our re-calibrated intrinsic \lya{} profile reconstructions. Our fitting pipeline is as follows:
\begin{enumerate}
\item We draw $\sim10^5$ intrinsic QSO profile estimates from the reconstructed \lya{} and \nv{} line profiles (see section~\ref{sec:Reconstruction}).
\item We convolve these $\sim10^5$ reconstructed profiles by the blueshift corrected template flux ratios to recalibrate for a systematic $\sim10$ per cent underestimate of the intrinsic flux as outlined in Section~\ref{sec:recal} leading to $\sim10^6$ corrected reconstruction profiles.
\item Each intrinsic profile is then multiplied by 10$^5$ synthetic damping wing opacities following Section~\ref{sec:Damping}. This results in $\sim10^{11}$ mock spectra for each \avenf\ snapshot from the EoR simulation.
\item Each $\sim10^{11}$ mock spectra is then individually compared to the observed QSO spectrum over the $\lambda=1218-1230$~\AA\ region (consistent with \citealt{Greig:2017,Greig:2019,Greig:2022}). For each, we calculate a $\chi^{2}$ relative to observed flux and the error spectrum. Where appropriate, absorption features or other problematic regions of the observed spectrum over this fitting range are masked out following visual inspection that could otherwise incorrectly bias our results.
\item We then obtain a likelihood for the current \avenf\ by averaging (marginalising) over all $\sim10^{11}$ mock spectra drawn from the corresponding \avenf\ snapshot.
\item Steps (ii)--(v) are then repeated for all available \avenf\ snapshots (24) to obtain a final 1D probability distribution function (PDF) of the \avenf\ for our particular observed QSO.
\end{enumerate}

Importantly, unlike in our previous works, in step (iv) we re-bin the observed spectrum onto 1\AA\ bins over the entire $\lambda=1218-1230$~\AA\ region. Additionally, we were also required to enlarge the amplitude of the provided flux errors by a factor of 5\footnote{We explored increasing this error by factors of 2, 5 and 10 and found no discernible difference in the inferred constraints other than a broadened PDF for increasing error. Thus we chose a factor of two to limit the amplitude of the increased error spectrum.}. These steps were necessitated owing to the considerably higher resolution and correspondingly small error spectrum provided by X-shooter. In the absence of this re-binning and enlarged errors, we found our simple analytic $\chi^{2}$ estimate of the likelihood would encounter numerical difficulties owing to sampling large $\chi^{2}$ values and correspondingly being too sensitive to narrow subsets of individual profiles who matched the observed data. Alternatively, we could have considered a narrow wavelength range, however, doing so is more prone to biasing the results to particular features in the data. This highlights the necessity to improve our joint fitting pipeline which we will pursue in future work, where we will move away from using analytic likelihoods altogether in favour of likelihood-free or simulation based inference (see e.g \citealt{Cranmer:2020} for a recent review on such methods and a preliminary exploration of simulation based inference for extracting constraints from QSO damping wings by \citealt{Chen:2023}).

\section{Results} \label{sec:Results}

\subsection{Reconstruction of the XQR-30 sample}

\begin{figure*}
  \centering
	\includegraphics[trim = 1cm 0.38cm 0cm 1cm, scale = 0.39]{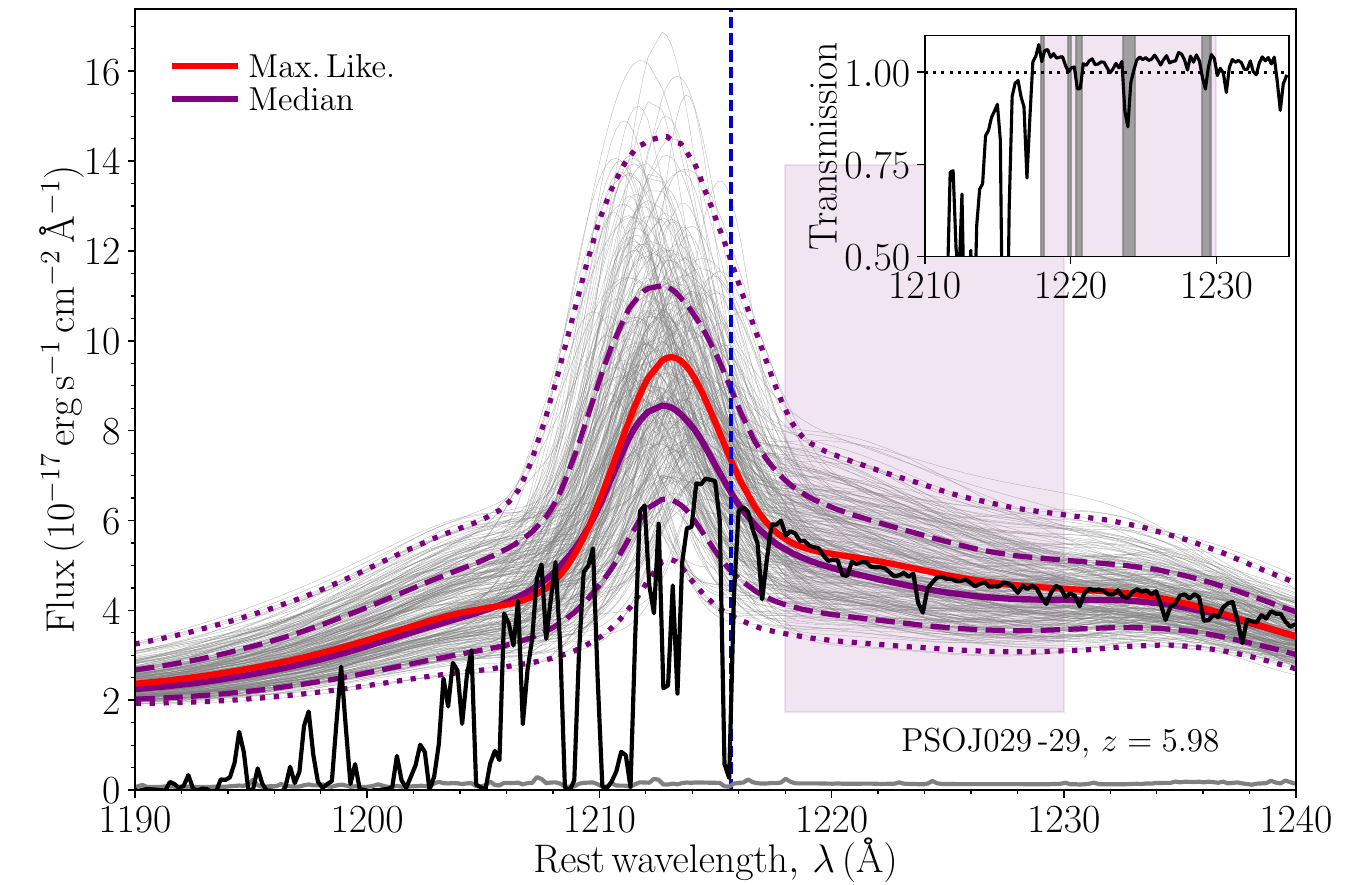}
	\includegraphics[trim = 0.5cm 0.38cm 1cm 1cm, scale = 0.39]{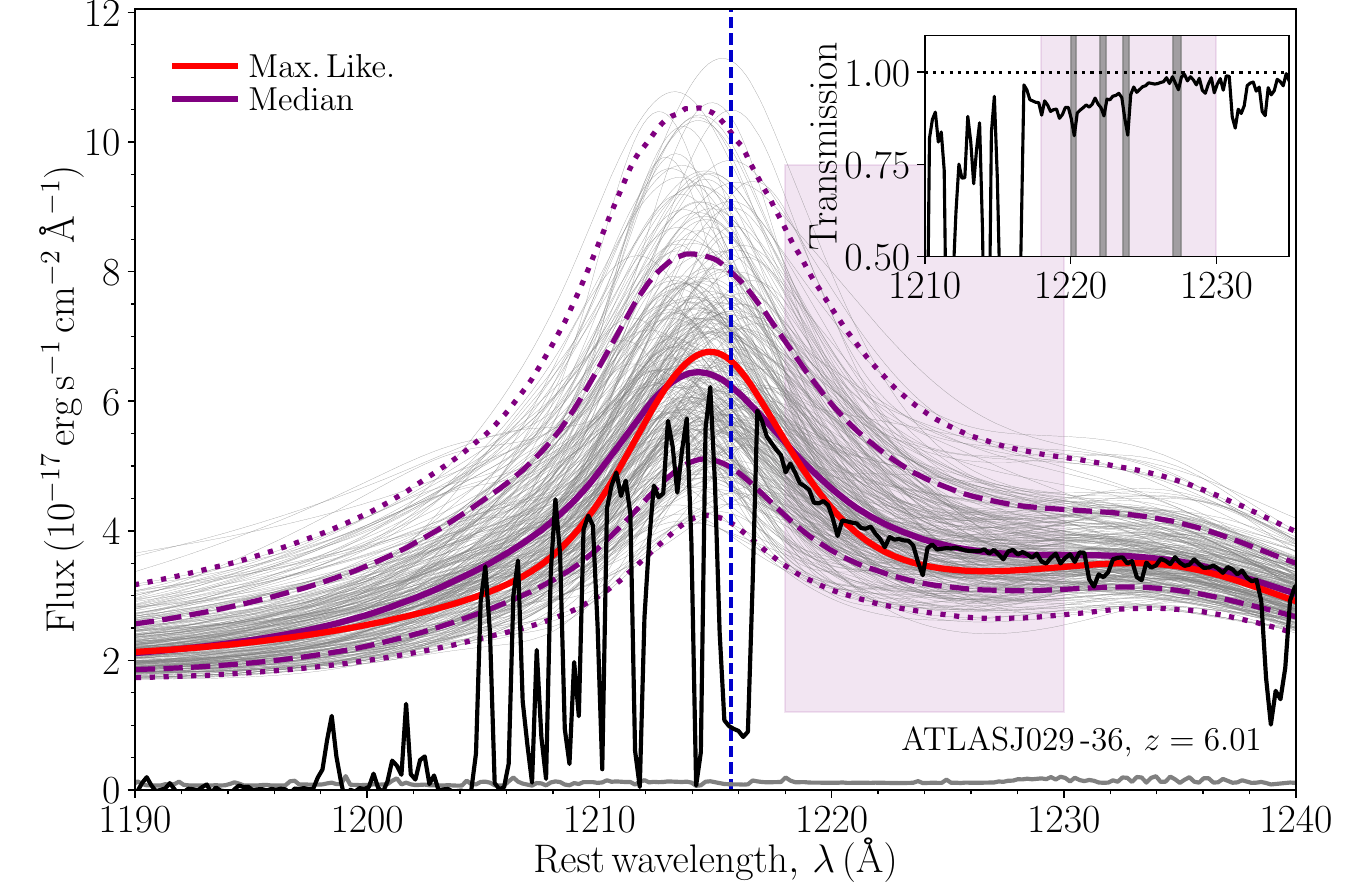}
	\includegraphics[trim = 1cm 0.38cm 0cm 0.4cm, scale = 0.39]{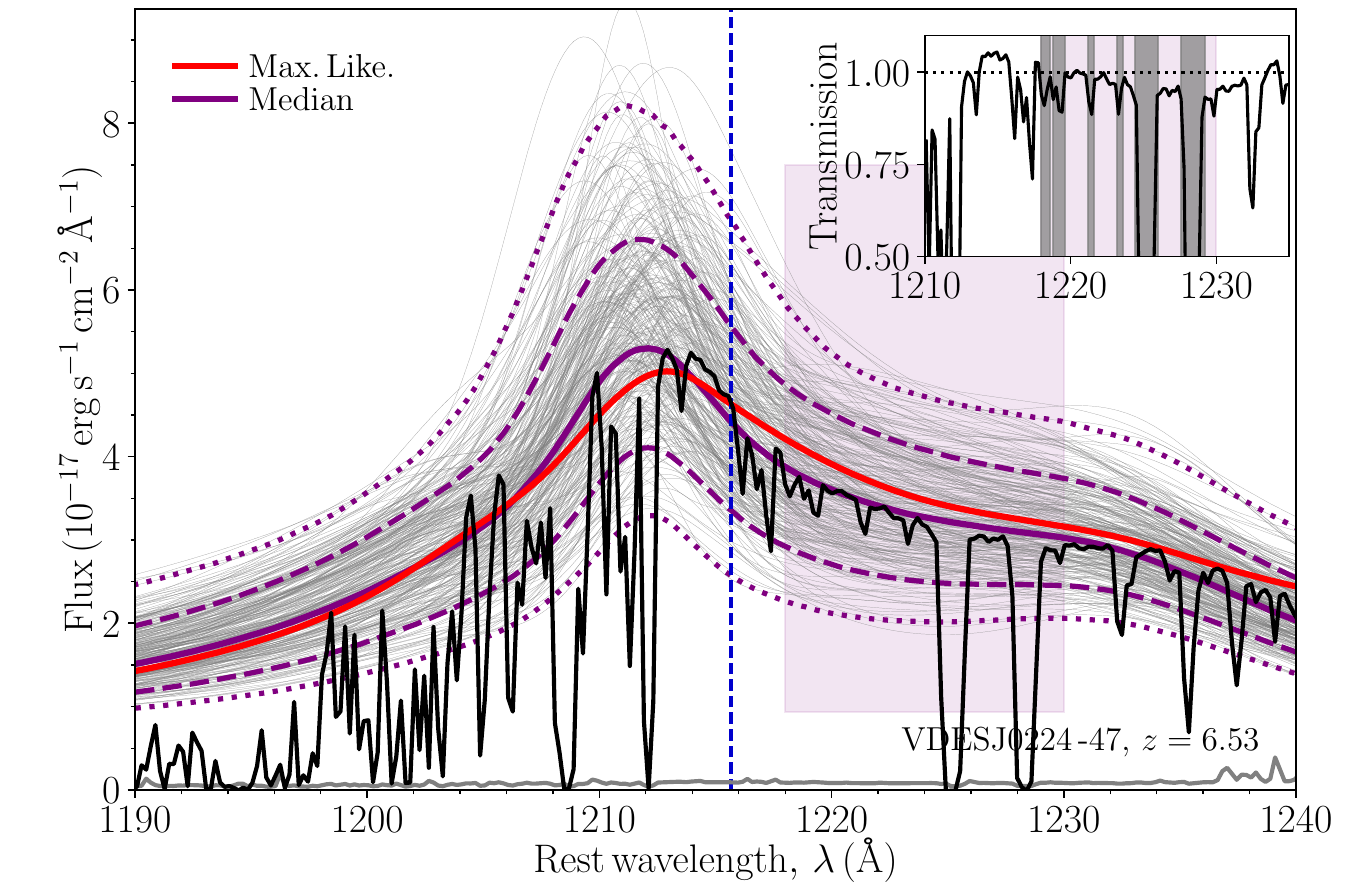}
	\includegraphics[trim = 0.5cm 0.38cm 1cm 0.4cm, scale = 0.39]{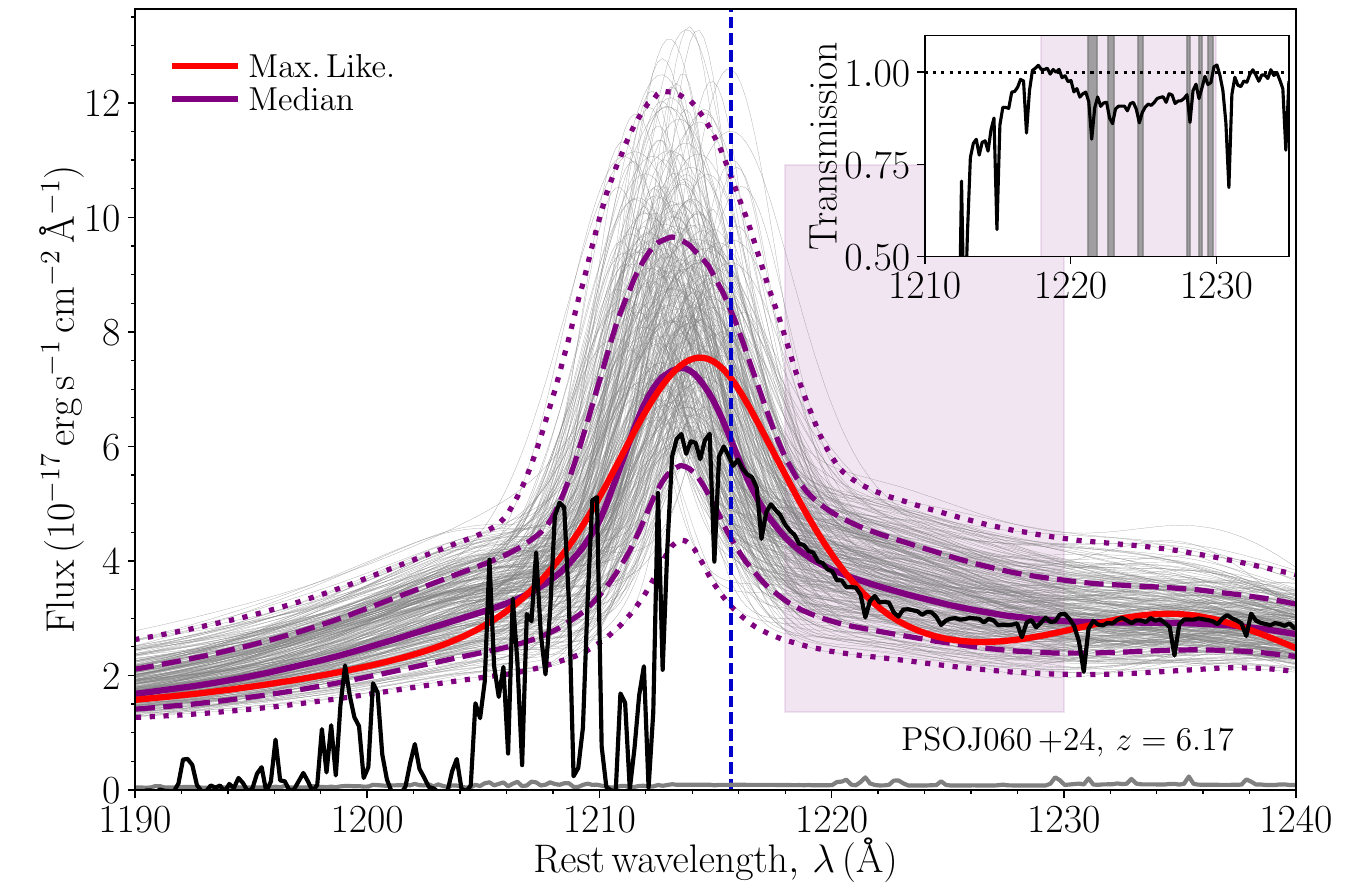}
	\includegraphics[trim = 1cm 0.4cm 0cm 0.4cm, scale = 0.39]{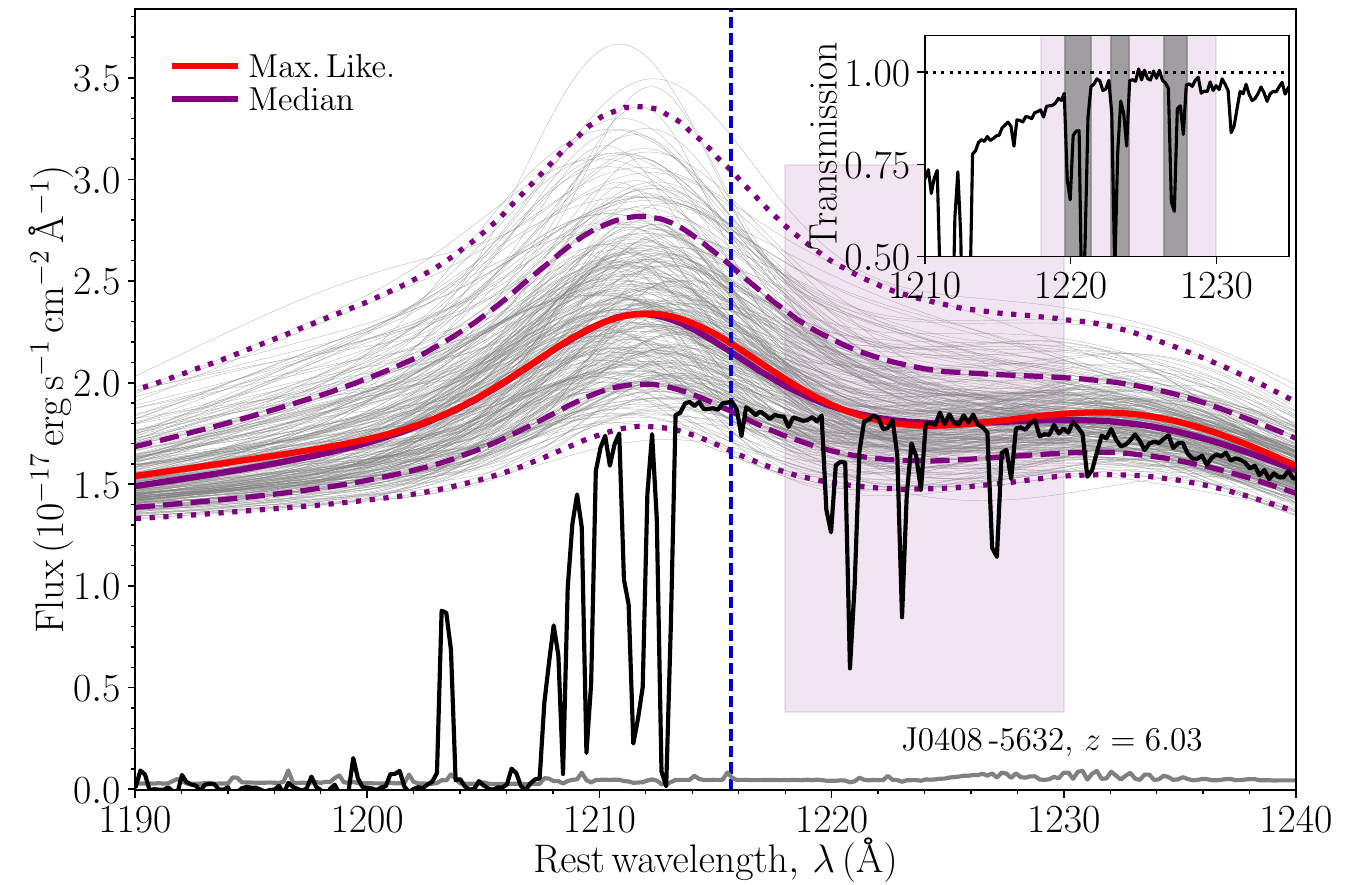}
	\includegraphics[trim = 0.5cm 0.4cm 1cm 0.4cm, scale = 0.39]{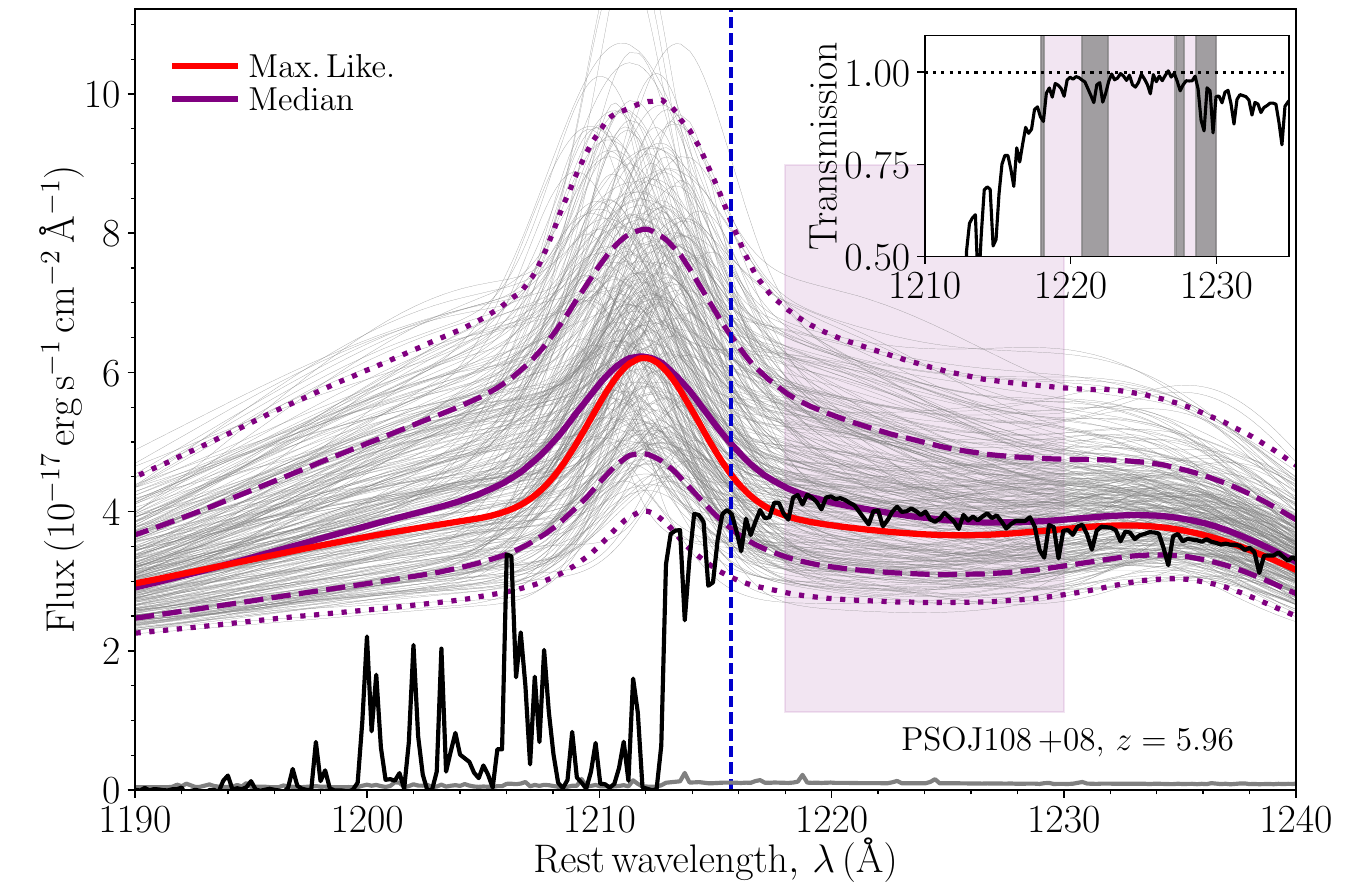}
	\includegraphics[trim = 1cm 0.5cm 0cm 0.4cm, scale = 0.39]{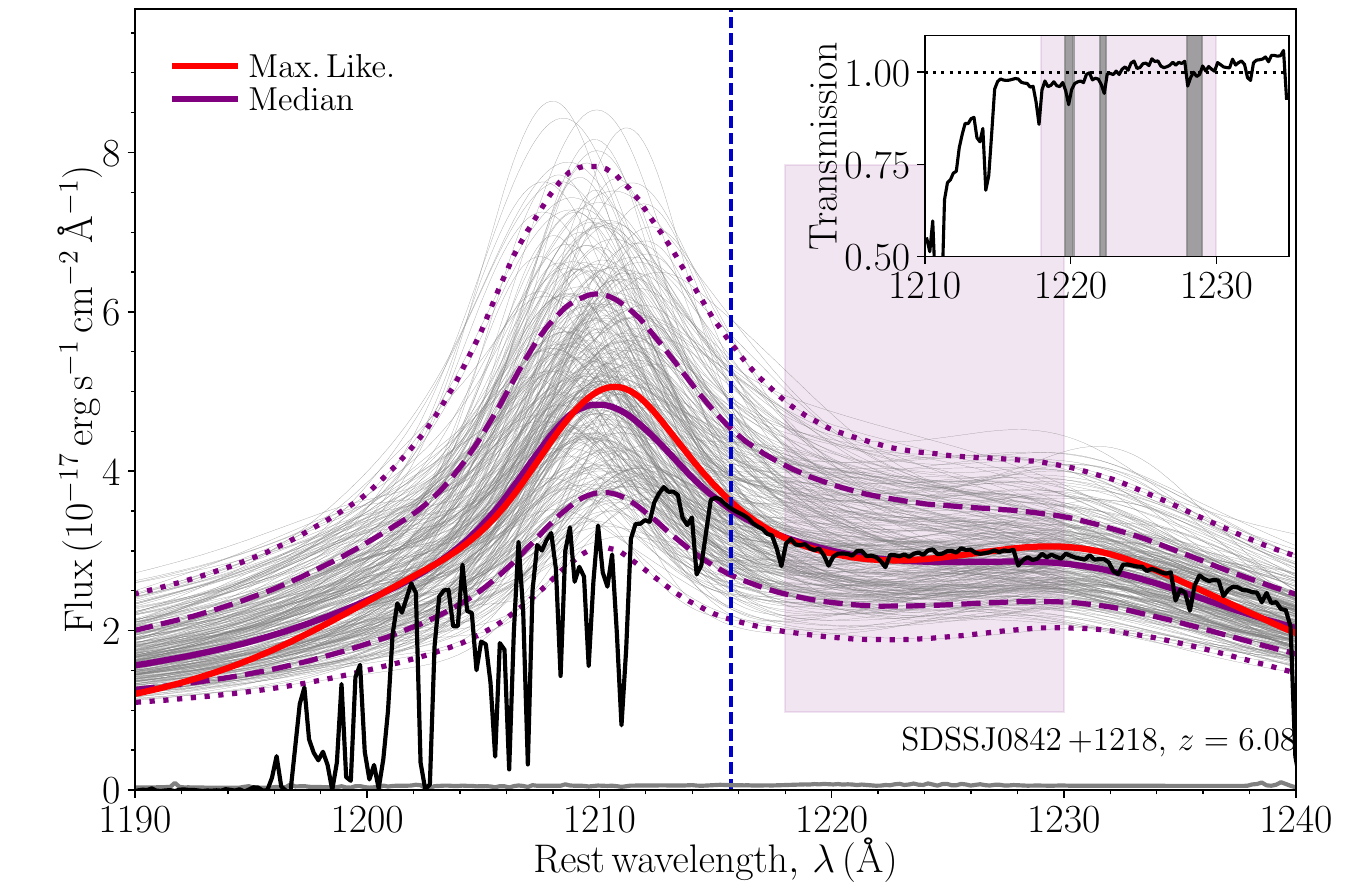}
	\includegraphics[trim = 0.5cm 0.5cm 1cm 0.4cm, scale = 0.39]{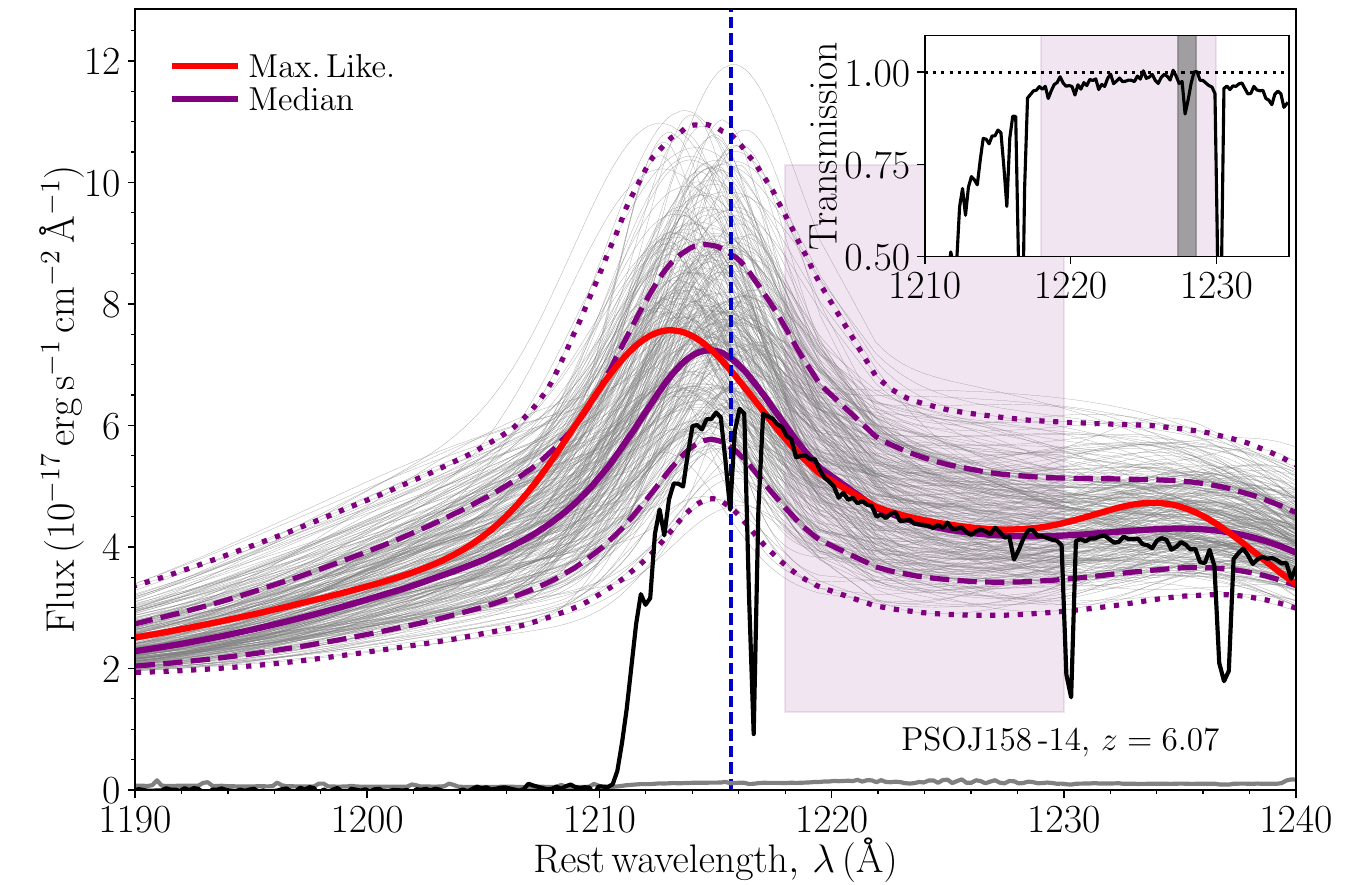}	  	  
   \caption{Intrinsic QSO reconstructions of the XQR-30 sample. We provide the maximum-likelihood reconstruction (red curve), the purple solid, dashed and dotted curves correspond to the median, 68th and 95th percentile profiles and the thin grey curves represent 300 random draws from our full posterior distribution. The solid thick grey curve near zero corresponds to the QSO noise spectrum, amplified by a factor of 5 to improve the numerical performance of our neutral fraction inference approach (see text for further details). The purple shaded box demarcates our damping wing fitting region (1218-1230\AA\ region). \textit{Inset panel:} The \lya{} transmission profile highlighting the presence (if any) of an IGM damping wing imprint (i.e. observed transmission spectrum below unity) assuming the median (purple curve). Grey shaded regions correspond to features in the observed spectrum removed from our likelihood fitting.}
  \label{fig:results}
\end{figure*}

\begin{figure*}
   \ContinuedFloat
     \centering
	\includegraphics[trim = 1cm 0.38cm 0cm 1cm, scale = 0.39]{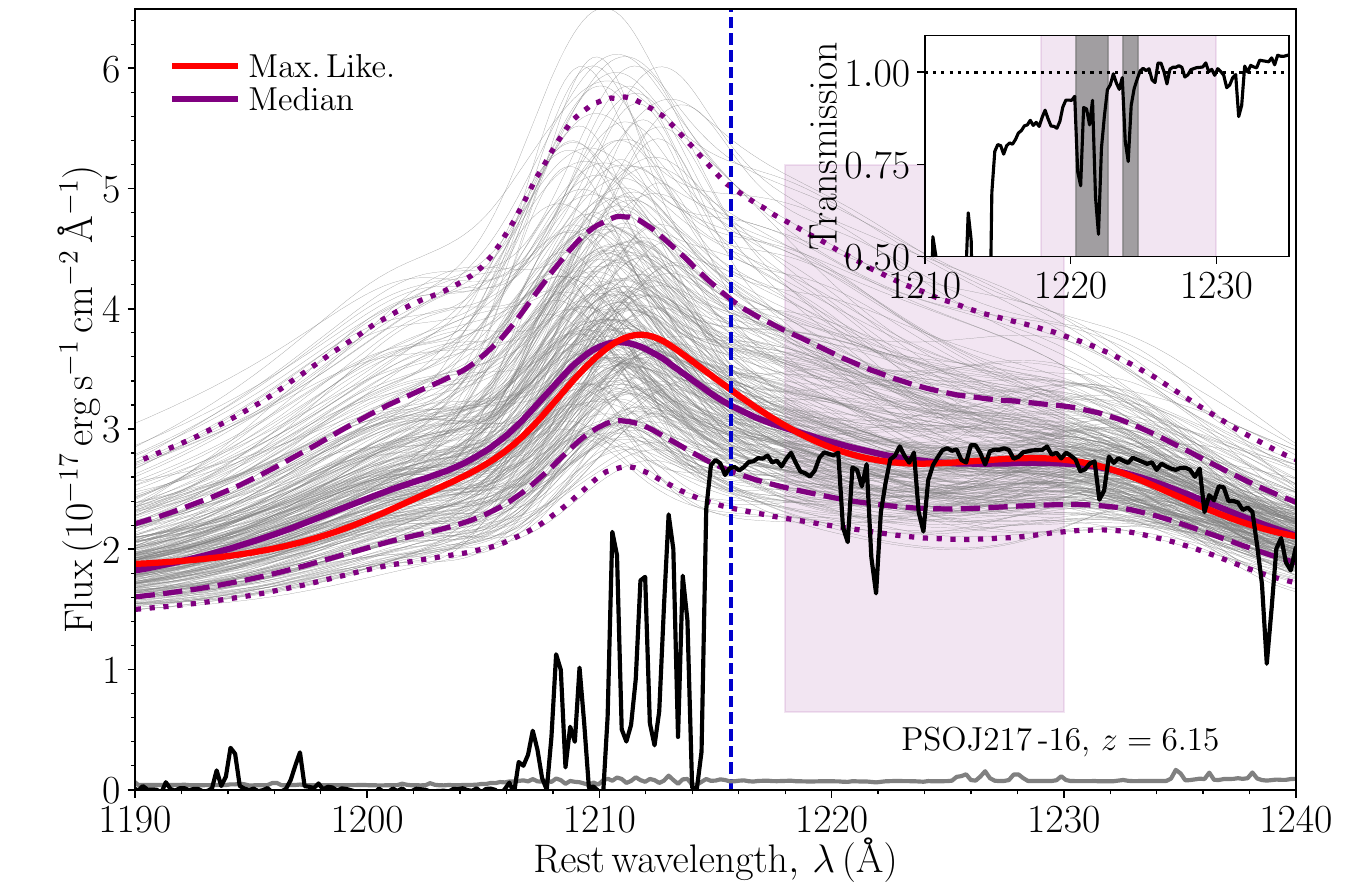}
	\includegraphics[trim = 0.5cm 0.38cm 1cm 1cm, scale = 0.39]{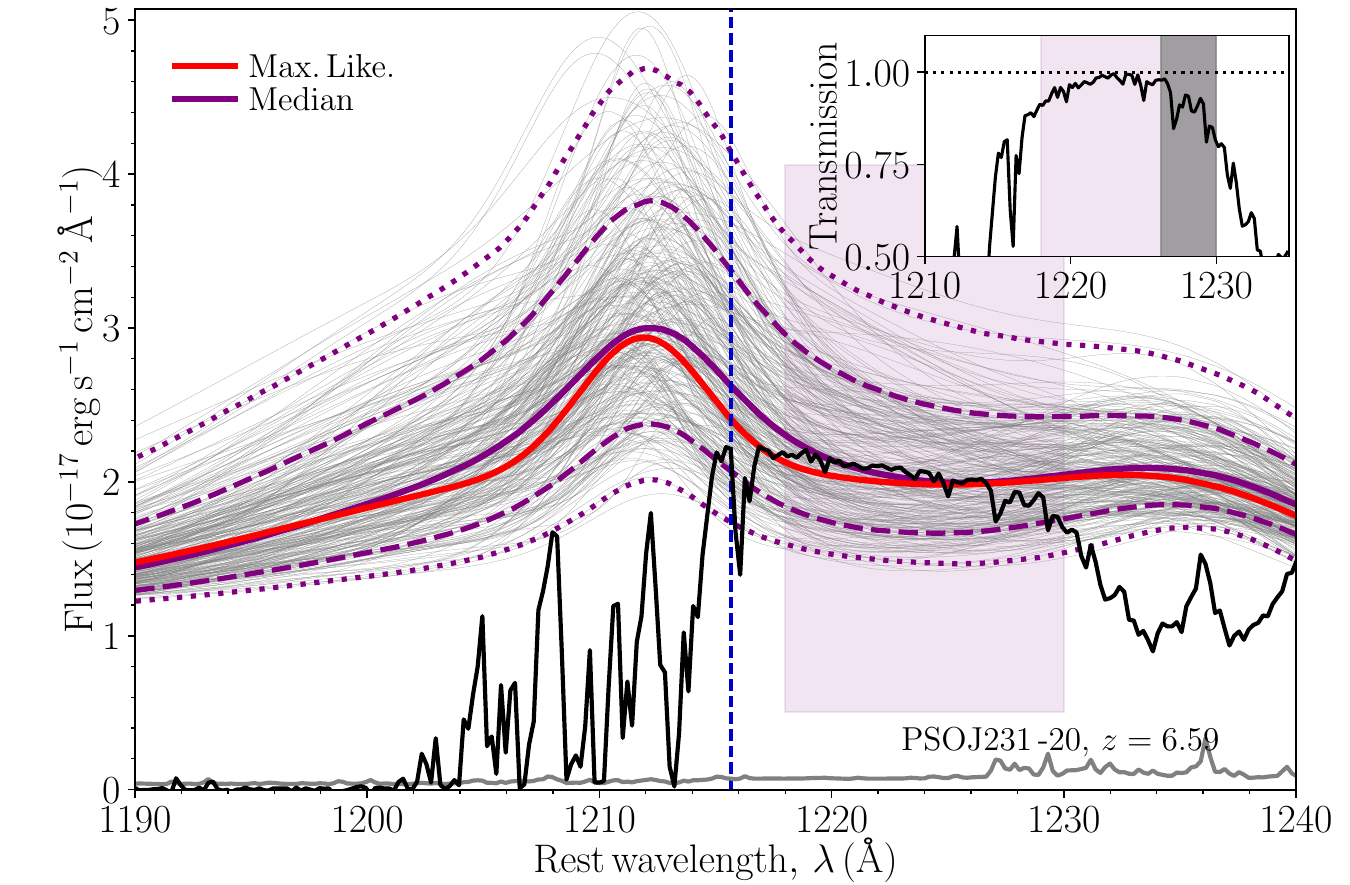}
	\includegraphics[trim = 1cm 0.38cm 0cm 0.4cm, scale = 0.39]{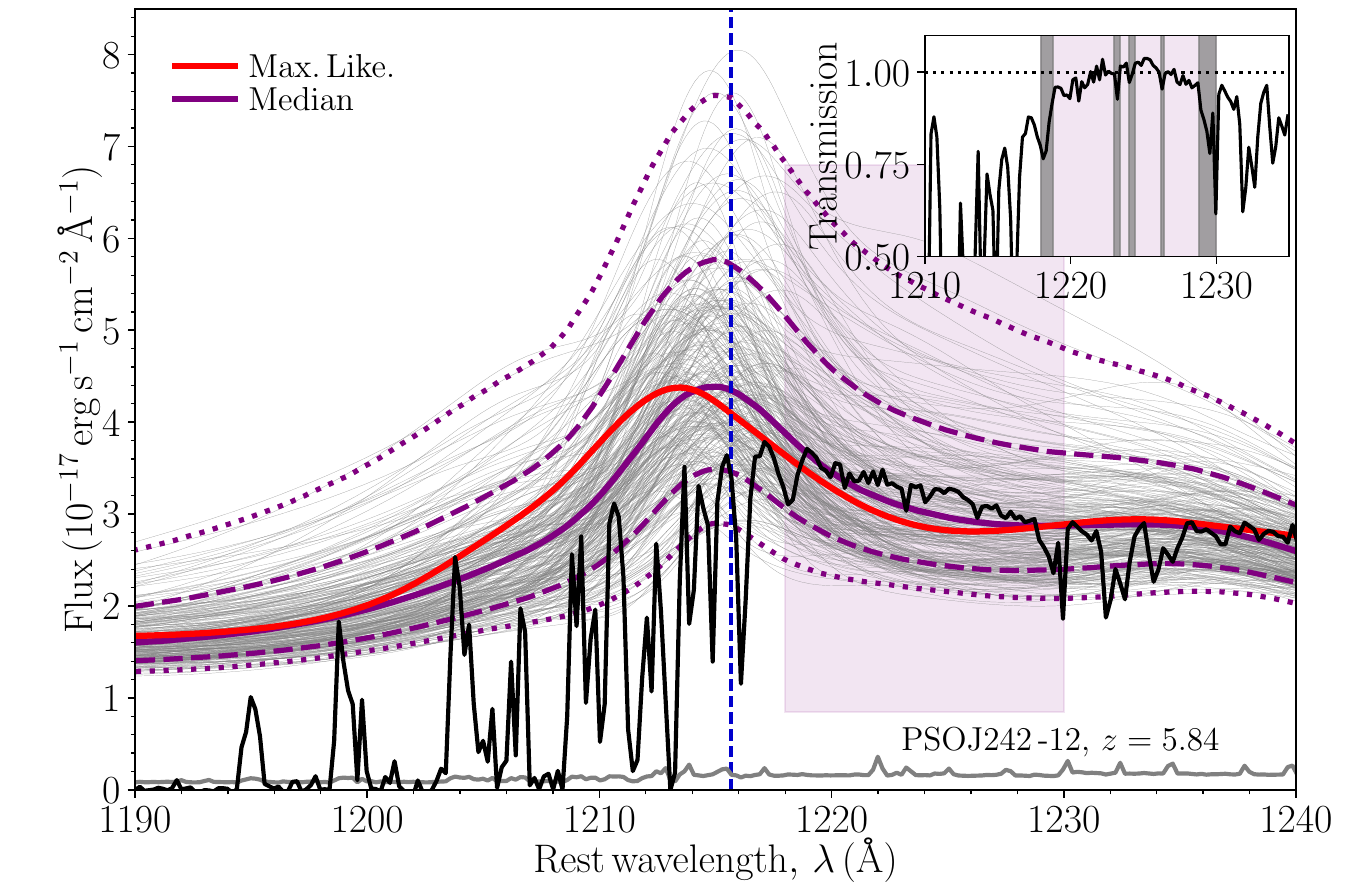}
	\includegraphics[trim = 0.5cm 0.38cm 1cm 0.4cm, scale = 0.39]{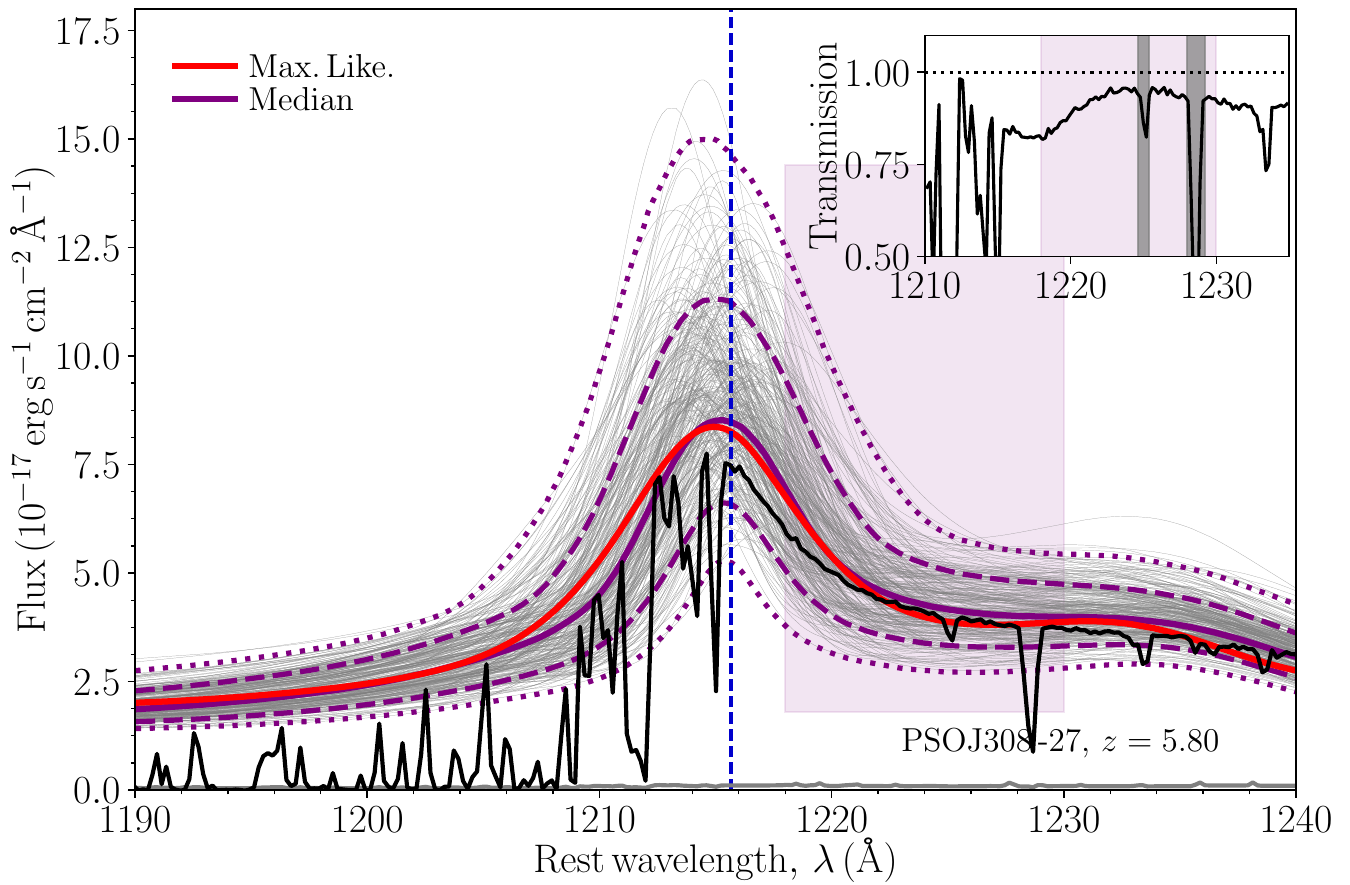}
	\includegraphics[trim = 1cm 0.4cm 0cm 0.4cm, scale = 0.39]{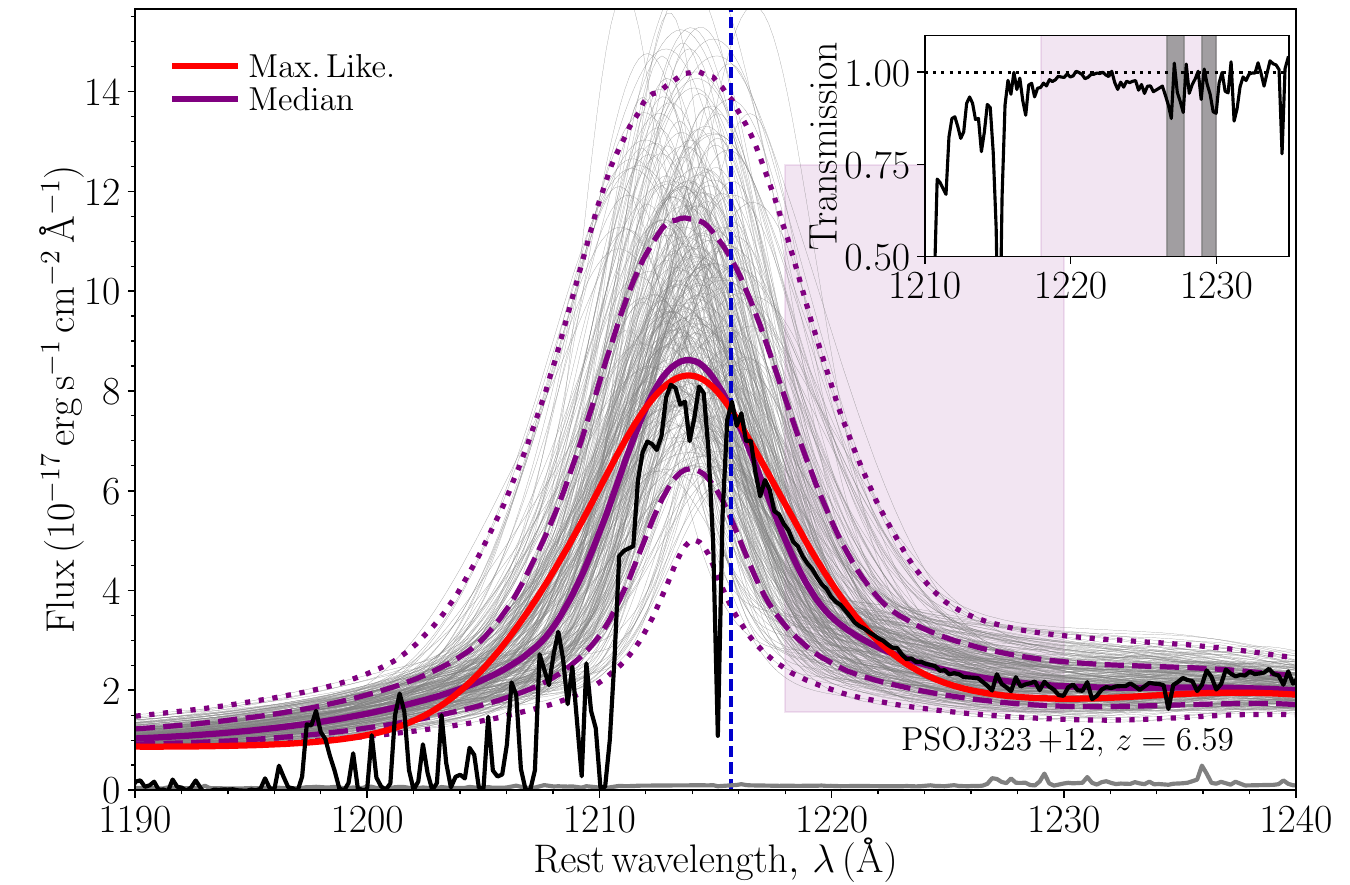}
	\includegraphics[trim = 0.5cm 0.4cm 1cm 0.4cm, scale = 0.39]{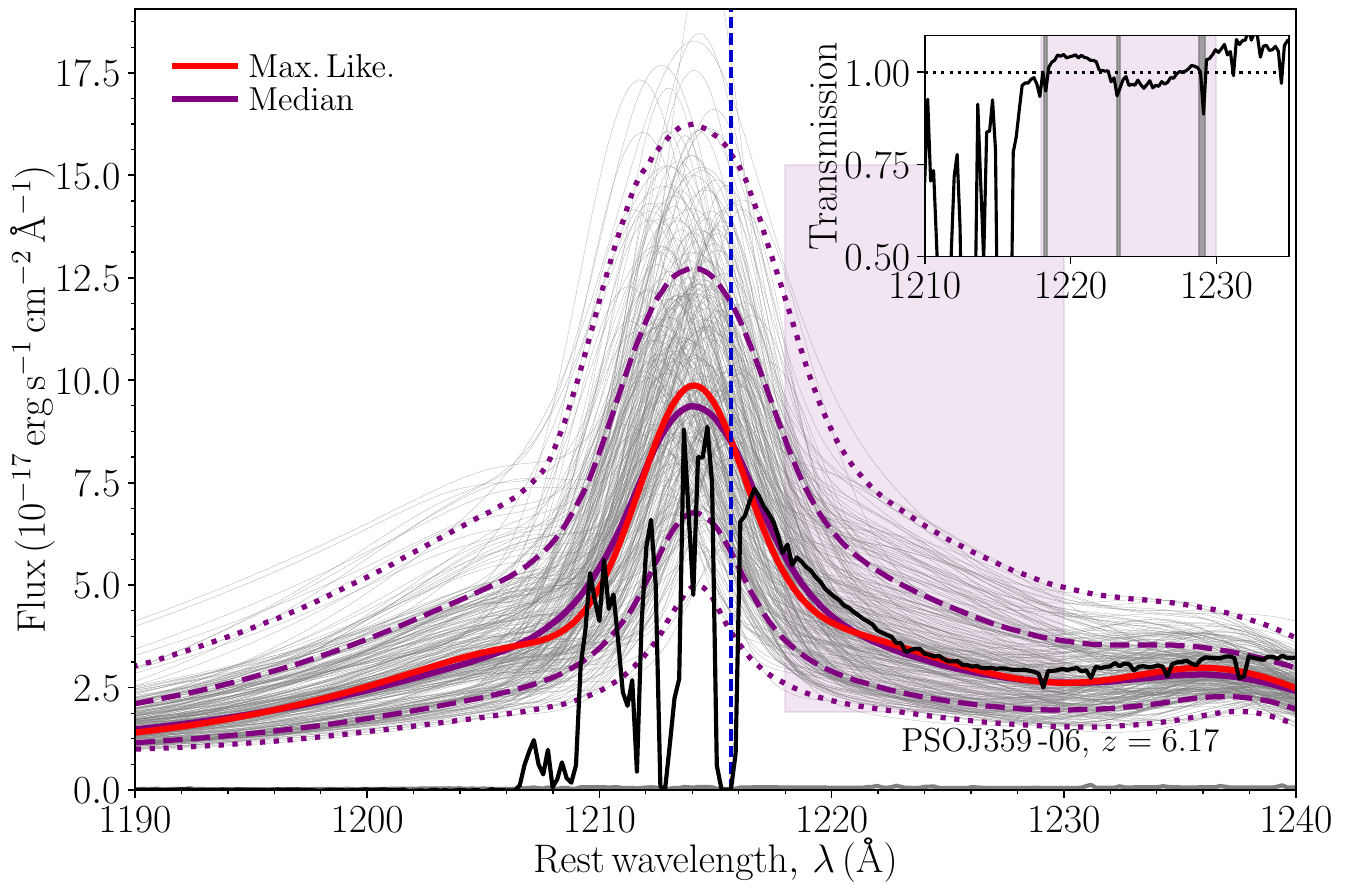}
	\includegraphics[trim = 1cm 0.5cm 0cm 0.4cm, scale = 0.39]{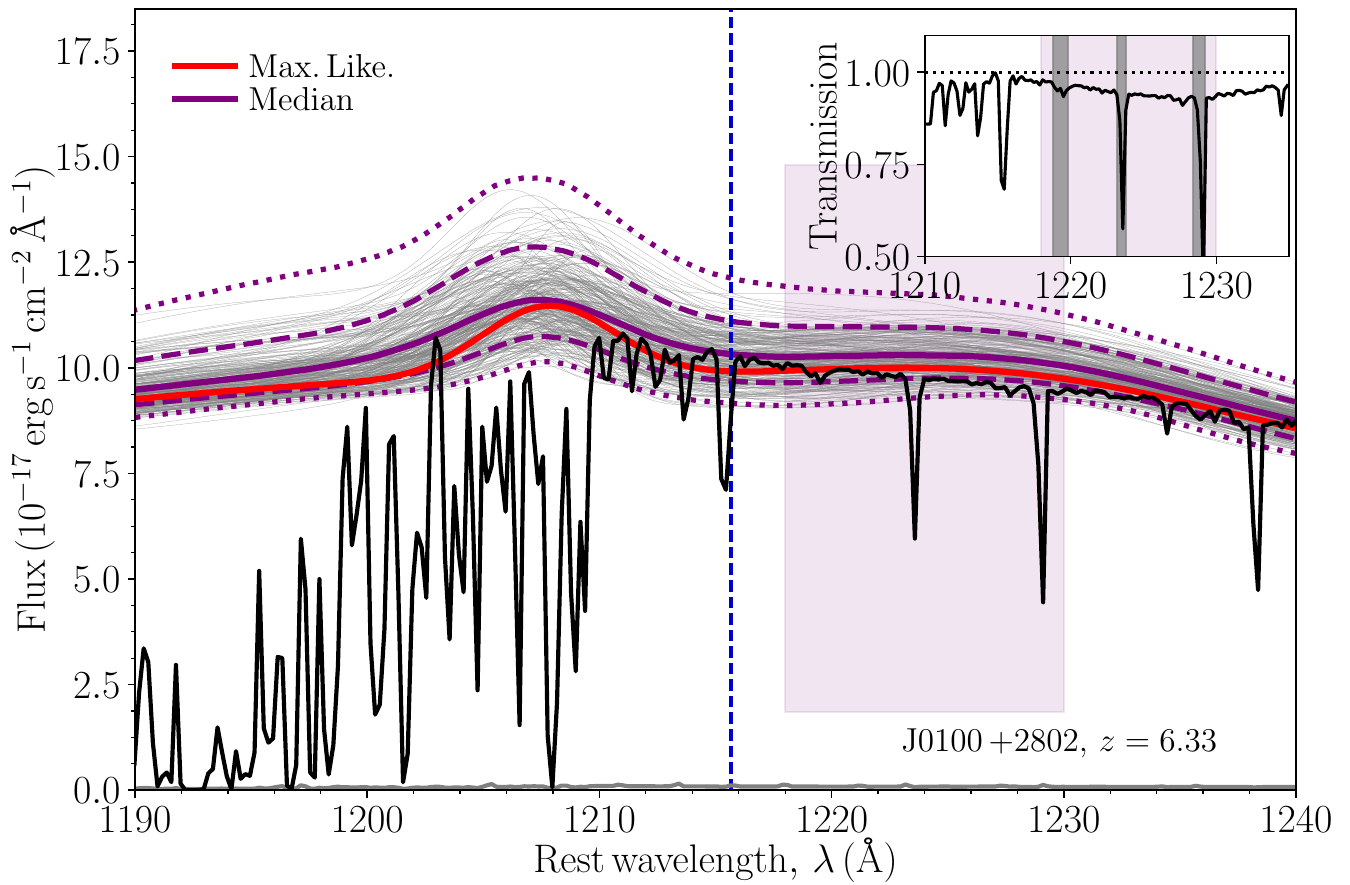}
	\includegraphics[trim = 0.5cm 0.5cm 1cm 0.4cm, scale = 0.39]{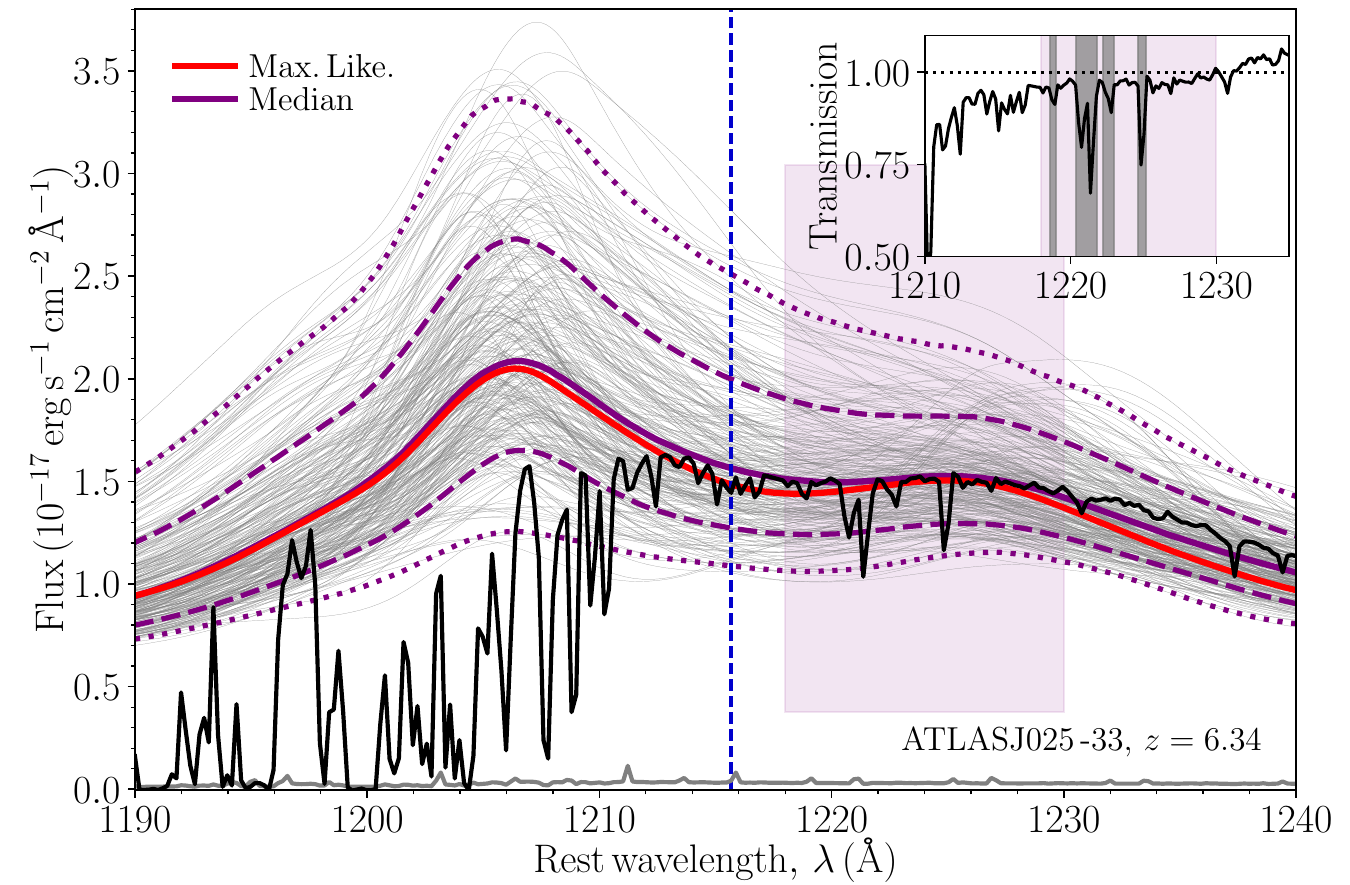}	  	  
   \caption{$-$\textit{continued}}
  \label{fig:results}
\end{figure*}

\begin{figure*}
   \ContinuedFloat
     \centering
	\includegraphics[trim = 1cm 0.38cm 0cm 1cm, scale = 0.39]{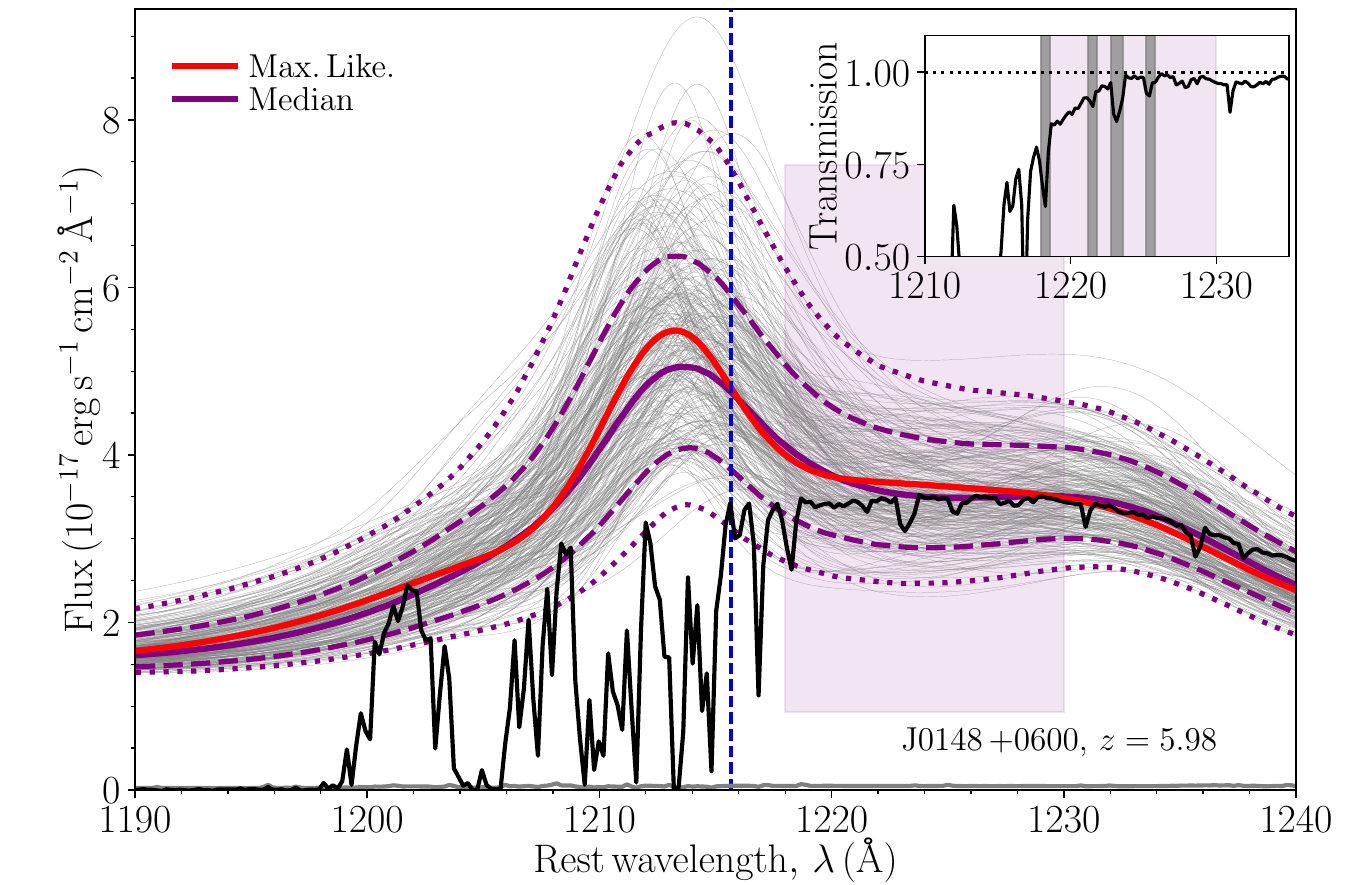}
	\includegraphics[trim = 0.5cm 0.38cm 1cm 1cm, scale = 0.39]{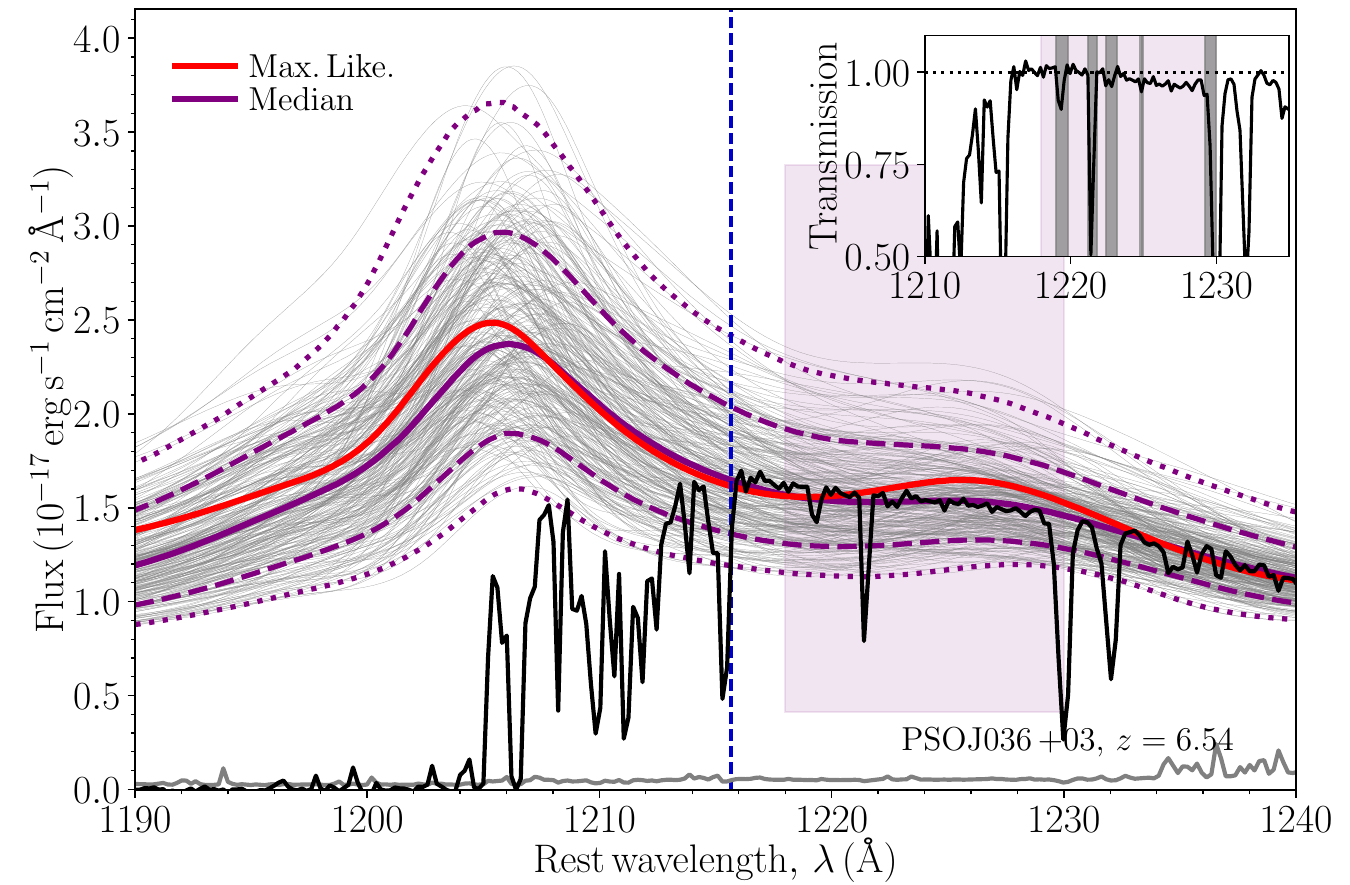}
	\includegraphics[trim = 1cm 0.38cm 0cm 0.4cm, scale = 0.39]{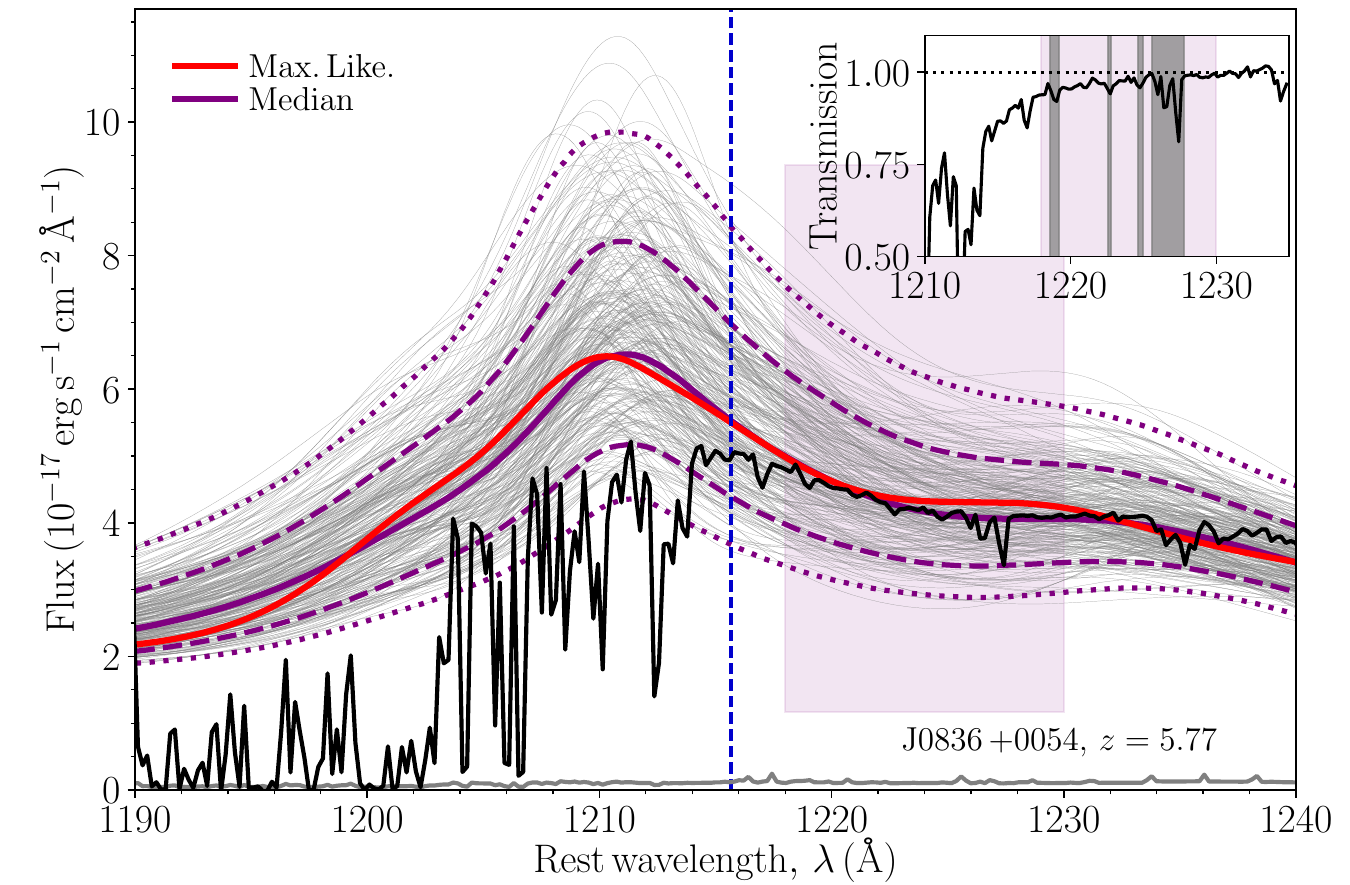}
	\includegraphics[trim = 0.5cm 0.38cm 1cm 0.4cm, scale = 0.39]{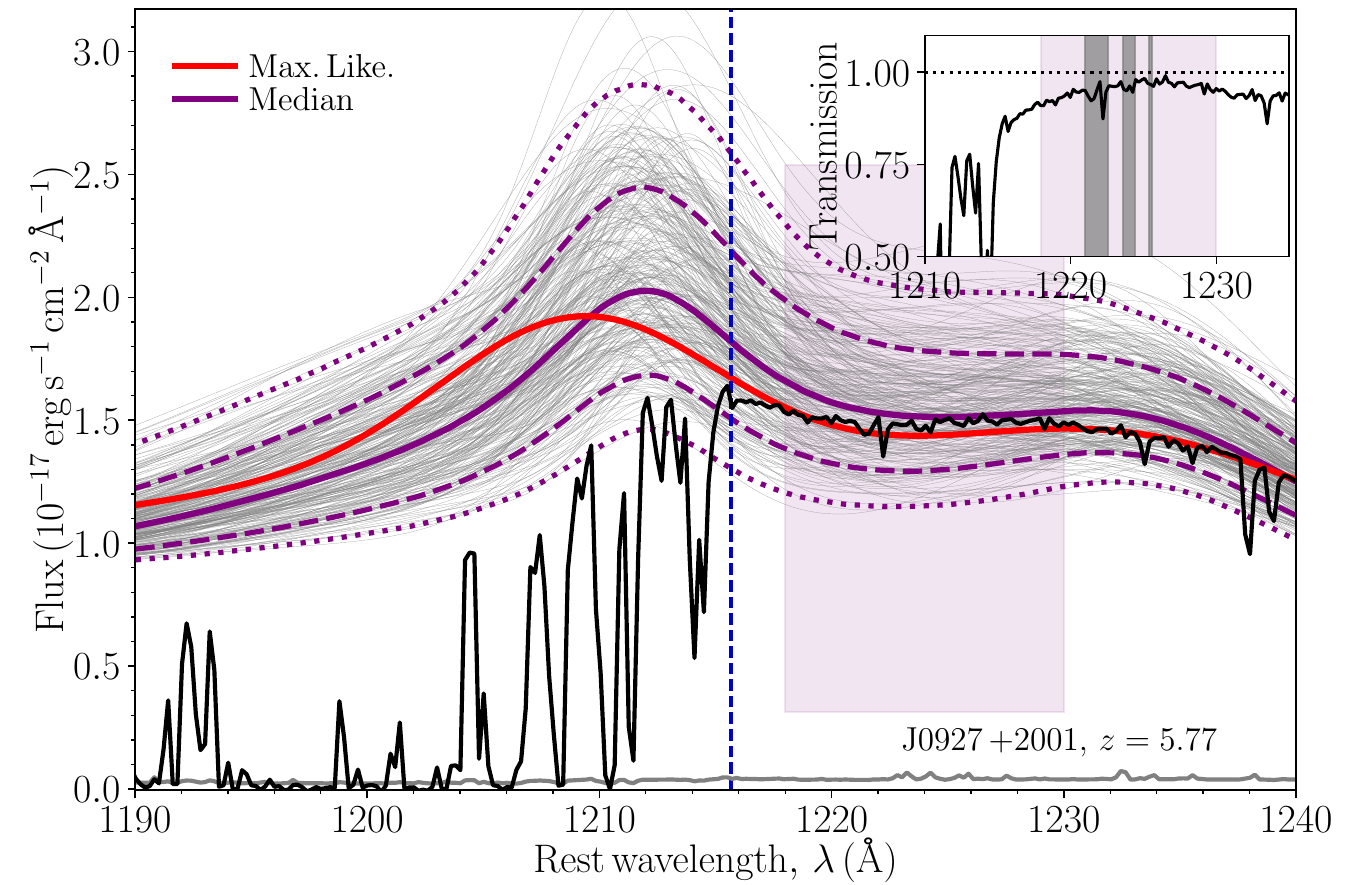}
	\includegraphics[trim = 1cm 0cm 0cm 0.4cm, scale = 0.39]{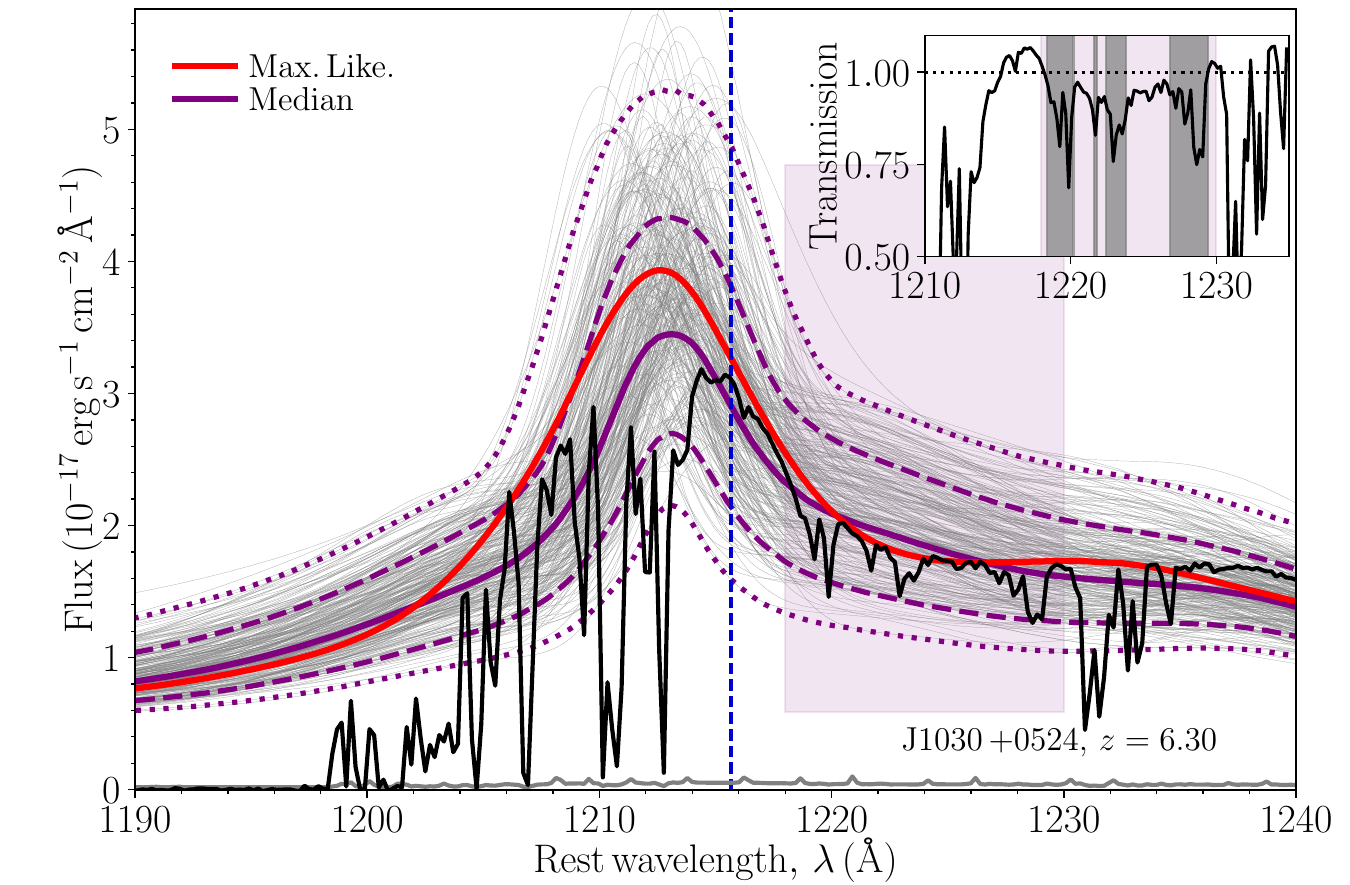}
	\includegraphics[trim = 0.5cm 0cm 1cm 0.4cm, scale = 0.39]{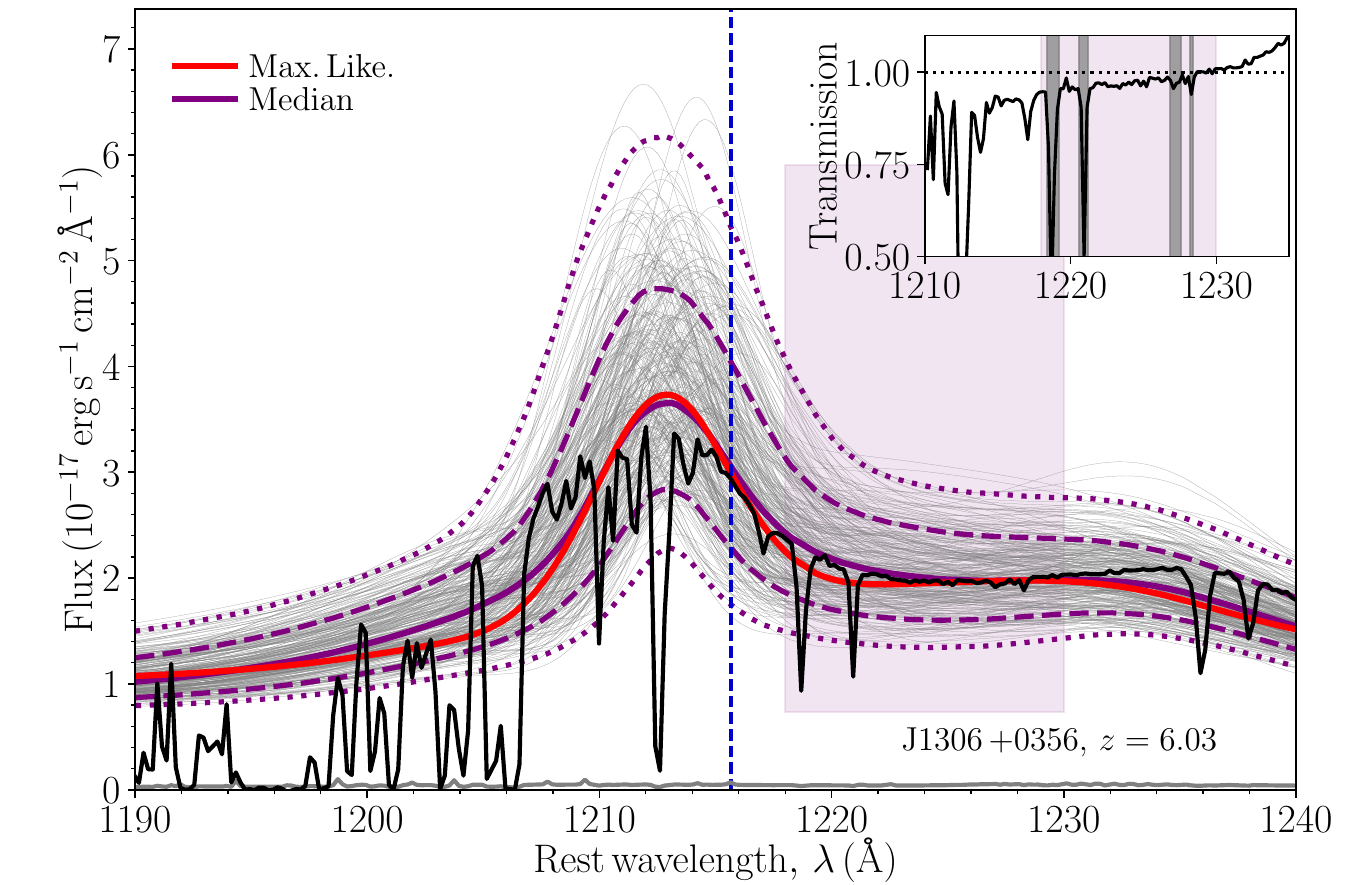}
	\includegraphics[trim = 1cm 0.4cm 0cm 0cm, scale = 0.39]{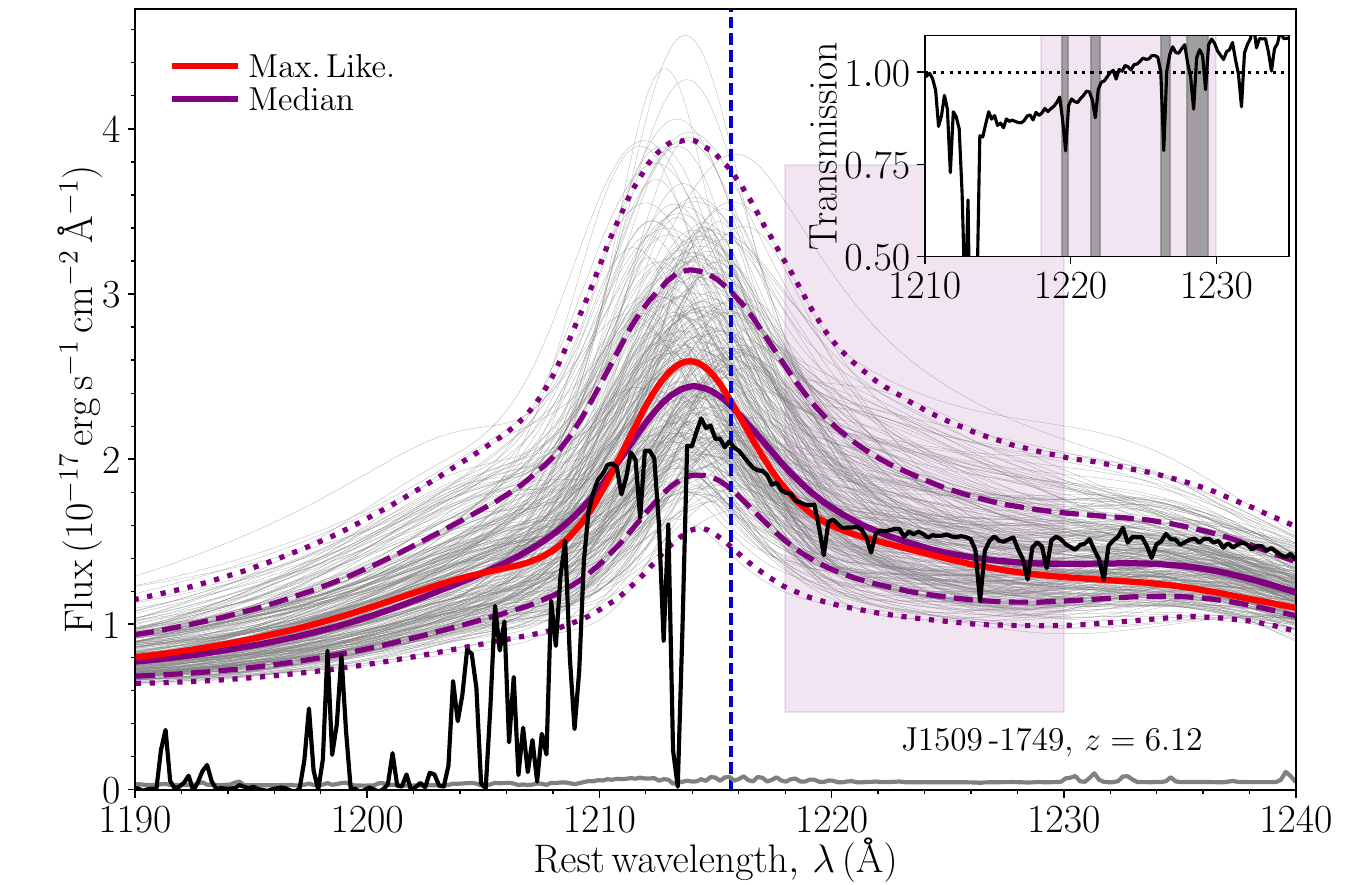}
   \caption{$-$\textit{continued}}
  \label{fig:results}
\end{figure*}

In Figure~\ref{fig:results} we provide the reconstructed intrinsic profiles for the 23 QSOs from the XQR-30 sample deemed suitable for IGM damping wing analysis (see Section~\ref{sec:sample}). In each panel we show the maximum-likelihood (ML) reconstruction profile (red curve), 300 random draws from our full posterior distribution (thin grey lines), the median (solid purple curve) and 68th (purple dashed) and 95th (purple dotted) percentile profiles obtained from the full posterior distribution. The purple shaded box corresponds to the 1218-1230\AA\ region over which we fit for the damping wing imprint. In the inset panel, we provide the \lya{} transmission profile over this 1218-1230\AA\ region obtained using the median profile, highlighting the presence (if any) of an IGM damping wing imprint. Grey shaded regions correspond to features removed from our likelihood fitting.

Here, we will limit our discussions of these reconstructed QSOs to focus only on broad observations that can be made across the sample, and only focus on individual QSOs that warrant further discussion. Firstly, for the most part, the \lya{} transmission profile within the 1218-1230\AA\ region either appears flat, consistent with little to no IGM attenuation, or monotonically increasing with rest-frame wavelength. This indicates that our recalibration step to our reconstruction pipeline is performing well in mitigating our observed systematic offset in the predicted intrinsic flux between the \lya{} and \nv{} emission lines (see Sections~\ref{sec:init} and~\ref{sec:recal}). 

However, potentially problematic QSO reconstructions remain, for example in the case of J060+24 and J359-06, which exhibit smoothly varying features in their transmission profiles. For the latter, this behaviour should not impact the results as the relative change in amplitude of the transmission profile is fairly modest. For the former, the shape appears to be driven more so by a stronger blueshift in the \lya{} line profile leading to the uptick in transmission near 1218\AA. As a result this may result in a slight underestimate of the neutral fraction. ATLASJ029-36, VDESJ0224-47 and J1509-1749 also may exhibit similar features to those above but to a notably lesser extent. However, what is important is that since we draw from the full posterior distribution of reconstructed profiles that are convolved with the synthetic damping wing templates, profiles that better match the observed spectrum should result in a higher likelihood and thus dominate the constraining power. That is, using the full distribution of joint fits will be a better match than the median transmission profile that is provided for visualisation purposes only.

\subsection{Recovered IGM damping wing constraints}

\begin{figure*} 
	\begin{center}
	  \includegraphics[trim = 1.2cm 3.5cm 0cm 0.5cm, scale = 0.82]{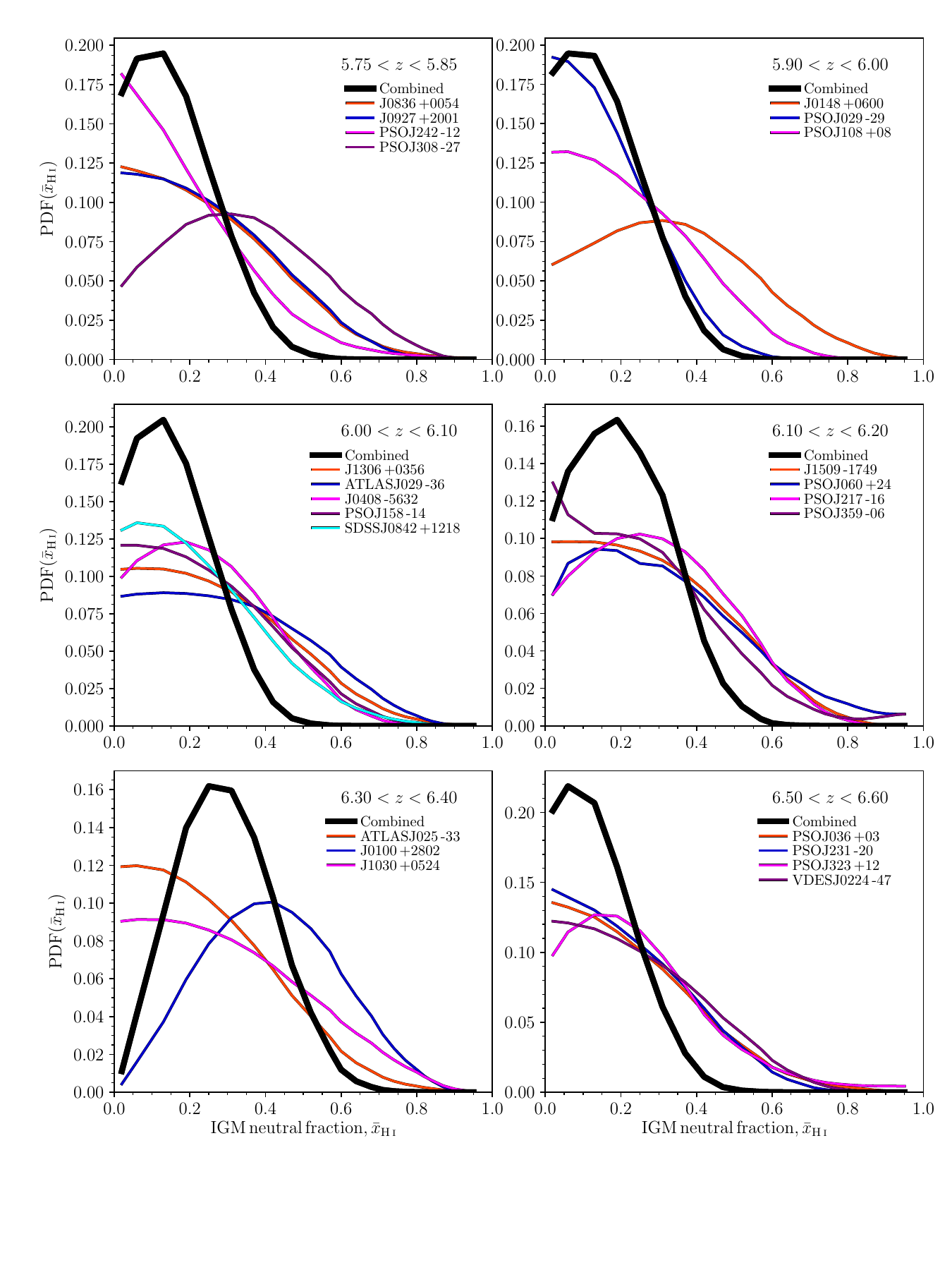}
	\end{center}
\caption[]{The marginalised 1D PDFs of the IGM neutral fraction for each of the XQR-30 QSOs, separated into redshift bins of $\Delta z = 0.1$. The thin coloured curves in each panel correspond to the recovered constraints for individual QSOs, whereas the thick black curve corresponds to the bin-averaged constraint on the IGM neutral fraction.}
\label{fig:BinnedPDFs}
\end{figure*}

In Figure~\ref{fig:BinnedPDFs}, we present the marginalised 1D PDFs of the IGM neutral fraction for each of the 23 QSOs of the XQR-30 sample. For this, we separate the QSOs into redshift bins of $\Delta z = 0.1$. In each panel, the thin coloured lines correspond to an individual QSO. For the vast majority of QSOs in the sample we recover broad, one-sided distributions, consistent with no IGM attenuation. This, however, is not too surprising given the relatively small decrement in the \lya{} transmission profiles shown in the inset panels of Figure~\ref{fig:results}. Nevertheless, we still find several QSOs with marginalised 1D PDFs consistent with a detection of IGM attenuation.

We summarise the individual constraints/limits on the IGM neutral fraction from the XQR-30 sample in the fourth column of Table~\ref{tab:XQR-30}. For marginalised 1D PDFs consistent with a detection, we present the median and 68th percentiles of the posteriors. For the 1D PDFs consistent with no attenuation, we present the 68th percentile upper limits, which is the IGM neutral fraction which encloses 68 per cent of the total probability.

\begin{figure*} 
	\begin{center}
          \includegraphics[trim = 0.1cm 0.5cm 0cm 0.5cm, scale = 0.7]{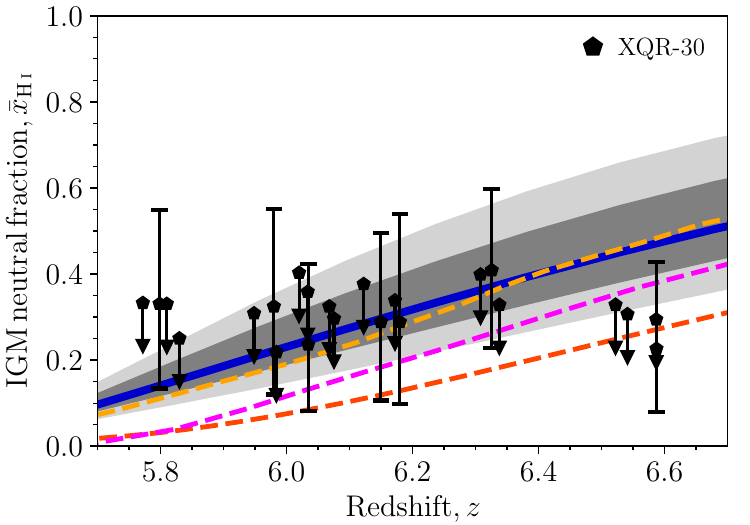}
          \includegraphics[trim = 0.4cm 0.5cm 0cm 0.5cm, scale = 0.7]{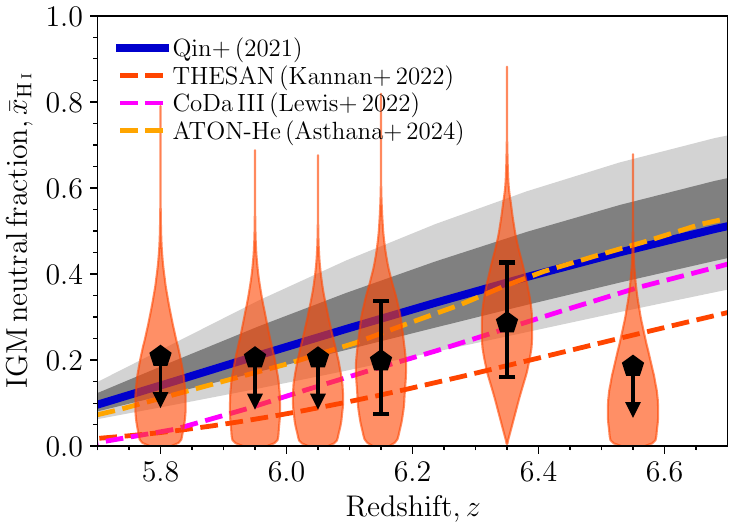}
	\end{center}
\caption[]{Our constraints on the IGM neutral fraction following our damping wing analysis of the XQR-30 sample. \textit{Left panel:} The individual constraints on the IGM neutral fraction. We present the median and 68th percentiles for our constraints and 68th percentile upper-limits for QSOs consistent with no IGM attenuation. \textit{Right panel:} A violin plot demonstrating the constraints on the IGM neutral fraction after binning our QSOs into $\Delta z = 0.1$ redshift bins (whereby the individually recovered PDFs of all QSOs within each redshift bin are multiplied to obtain an averaged constraint). In both panels, the blue curve and the dark and light shaded regions correspond to the median, 68th and 95th percentile EoR histories obtained by \citet{Qin:2021} (see text for further details). Further, the dashed red, magenta and orange curves correspond to the fiducial reionisation histories of state-of-the-art numerical simulations of the IGM from THESAN \citep{Kannan:2022,Garaldi:2022}, CoDA III \citep{Lewis:2022} and \citet{Asthana:2024}.}
\label{fig:BinnedHistory_Individual}
\end{figure*}

In the left panel of Figure~\ref{fig:BinnedHistory_Individual}, we plot our individual QSO constraints on the IGM neutral fraction as a function of redshift. In Table~\ref{tab:binnedxHI} we summarise our joint constraints on the IGM neutral fraction after binning in redshift. In the right panel of Figure~\ref{fig:BinnedHistory_Individual} we provide the violin plots of the IGM neutral fraction after binning our individual QSOs into separate $\Delta z = 0.1$ redshift bins based loosely on the natural redshift sampling of the available QSOs in the XQR-30 sample. This choice of redshift sampling is fairly arbitrary but is adopted to ensure each bin contains multiple QSOs. We explored several alternative choices in binning (i.e. larger), but found the results to be generally insensitive to the choice owing to the natural redshift sampling of the data. 

In both panels we also provide the constrained EoR histories from the state-of-the-art Monte-Carlo Markov Chain (MCMC) analysis of \citet[][]{Qin:2021}. Here, these authors ran semi-numerical simulations of the 21-cm signal during the EoR using \cmfst{} \citep[][]{Mesinger:2007p122,Mesinger:2011p1123,Murray:2020} coupled with a hybrid scheme to model the \lya{} forest. Using these simulations, their astrophysical model describing the galaxies responsible for reionisation are constrained only against existing observational constraints on the reionisation epoch such as; the observed UV galaxy LFs at $z=6-10$, the electron scattering optical depth, $\tau_{\rm e}$, measured by Planck \citep{Planck:2018}, the dark pixel limits on the IGM neutral fraction \citep{McGreer:2015p3668} and PDFs of the \lya{} effective optical depth from the \lya{} forest at $z=5-6$ \citep{Bosman:2018}. Post-processing of the resultant posteriors on their inferred astrophysical model then yields a posterior on the allowed reionisation history given the existing observational data. We overlay the median (blue curve) and the 68th (dark grey) and 95th (light grey) percentile regions of the constrained EoR histories.

The vast majority of our inferred IGM damping wing constraints from the XQR-30 sample strongly align with the inferred EoR histories from \citet{Qin:2021}. Further, in general our XQR-30 data-points monotonically increase in IGM neutral fraction for increasing redshift, consistent with ongoing reionisation. The main discrepancy with our results and those from \citet{Qin:2021} are the inferred IGM neutral fraction limits at $z\gtrsim6.5$. Our IGM damping wing limits at $z\gtrsim6.5$ are systematically below the inferred EoR histories from \citet{Qin:2021}. Our one constraint within this bin, PSOJ323+12, returns an IGM neutral fraction of $\bar{x}_{\hi{}} = 0.23\substack{+0.19 \\ -0.15}$, which owing to the broad 68th percentiles is consistent with the inferred EoR history. For the remaining three QSOs, we recover upper-limits at $z\gtrsim6.5$ that are inconsistant at between 68th and 95th percentiles of the joint distribution. In Appendix~\ref{sec:prior} we explored the role of the assumed prior on our minimum \hii{} bubble size, $R_{\rm min}$. For our available 4 choices of $R_{\rm min}$, we observe a shift of $\sim0.02$ in the inferred IGM neutral fraction from our lowest to highest choice of $R_{\rm min}$ therefore our adopted choice of averaging over $R_{\rm min}$ is not likely a contributing factor to the lower IGM neutral fractions. Within the literature there are a broad range of approaches to both extract the damping wing imprint and to predict the intrinsic QSO profile, thus it would be prudent to perform a re-analyses of these QSOs using these alternative reconstruction/damping wing fitting pipelines to add further confidence to the results presented in this work. However, this is a significant undertaking and we leave such a re-analysis to future work.

To place these constraints into a broader context, we additionally provide the reionisation histories from recent state-of-the-art radiation-hydrodynamic simulations from THESAN \citep[][]{Kannan:2022,Garaldi:2022} and Cosmic Dawn III \citep[CoDa III][]{Lewis:2022}. We also provide the reionisation history from \citet{Asthana:2024} who post-processed the Sherwood simulation suite \citep{Bolton:2017} using radiative transfer. Note that unlike the \citet{Qin:2021} posteriors, which are constrained specifically by the \citet{Bosman:2018} \lya{} forest data, the THESAN and CoDA III simulations are single simulations designed to match a considerably broader range of existing observational data. As a result, their reionisation histories will differ to those from \citet{Qin:2021}. On the other hand, the \citet{Asthana:2024} simulations are specifically designed to match the XQR-30 \lya{} forest data of \citet{Bosman:2022} and as such the reionisation history is very similar to the median history recovered by \citet{Qin:2021}.

Our individual inferred IGM damping wing constraints form the XQR-30 sample remain broadly consistent with the reionisation histories from these state-of-the-art numerical simulations. Although the THESAN and CoDa III reionisation histories are $\Delta\bar{x}_{\hi{}} \sim0.1-0.15$ lower than that expected explicitly from the \lya{} forest, since the vast majority of our results are only upper limits these remain statistically consistent. For the handful of constraints that we recover, the CoDa III simulation is consistent with all but one constraint at $z\sim5.8$ whereas the THESAN simulations are inconsistent with several of our constraints. However, as noted above these simulations are not explicitly constrained to match the \lya{} forest data, but instead to broadly match a large range of existing observations. Thus any inconsistencies are not too concerning. Importantly, at $z\gtrsim6.5$ where our XQR-30 damping wing results begin to disagree with the inferred constraints from the \lya{} forest data, the THESAN and CoDA III simulations are perfectly consistent within the 68th percentile uncertainties of our constraints and upper-limits. Therefore the stronger inconsistencies with our damping wing results and the inferred constraints on the reionisation history using the \lya{} forest data at $z\sim6.5$ can likely be softened once additional observational constraints are folded into such an analysis. Nevertheless, once we bin the individual constraints (right panel of Figure~\ref{fig:BinnedHistory_Individual}) our inferred limits at $z\gtrsim6.5$ are still inconsistent at more than the 68th percentiles with all these numerical simulations.

\begin{table}
\caption{Summary of the redshift binned IGM neutral fraction constraints from our damping wing analysis. For IGM neutral fractions posteriors consistent with a detection, we provide the median and 68th percentiles. Remaining constraints are presented as 68th percentile upper-limits.}
\begin{center}
\begin{tabular}{@{}lccc}
\hline
Redshift range & $\bar{x}_{\hi{}}$ & QSOs in bin\\
\hline
\vspace{0.8mm}
$5.85 \le z < 5.95$ & $<0.21$ & 4 \\
$5.90 \le z < 6.00$ & $<0.20$ & 3 \\
$6.00 \le z < 6.10$ & $<0.21$ & 5 \\
$6.10 \le z < 6.20$ & $0.20\substack{+0.14 \\ -0.12}$ & 4 \\
$6.30 \le z < 6.40$ & $0.29\substack{+0.14 \\ -0.13}$ & 3 \\
$6.50 \le z < 6.60$ & $<0.18$ & 4 \\
\hline
\end{tabular}
\end{center}
\label{tab:binnedxHI}
\end{table}

\subsection{Compilation of reionisation constraints}

\begin{figure*} 
	\begin{center}
	  \includegraphics[trim = 0.5cm 0.8cm 0cm 0.5cm, scale = 0.75]{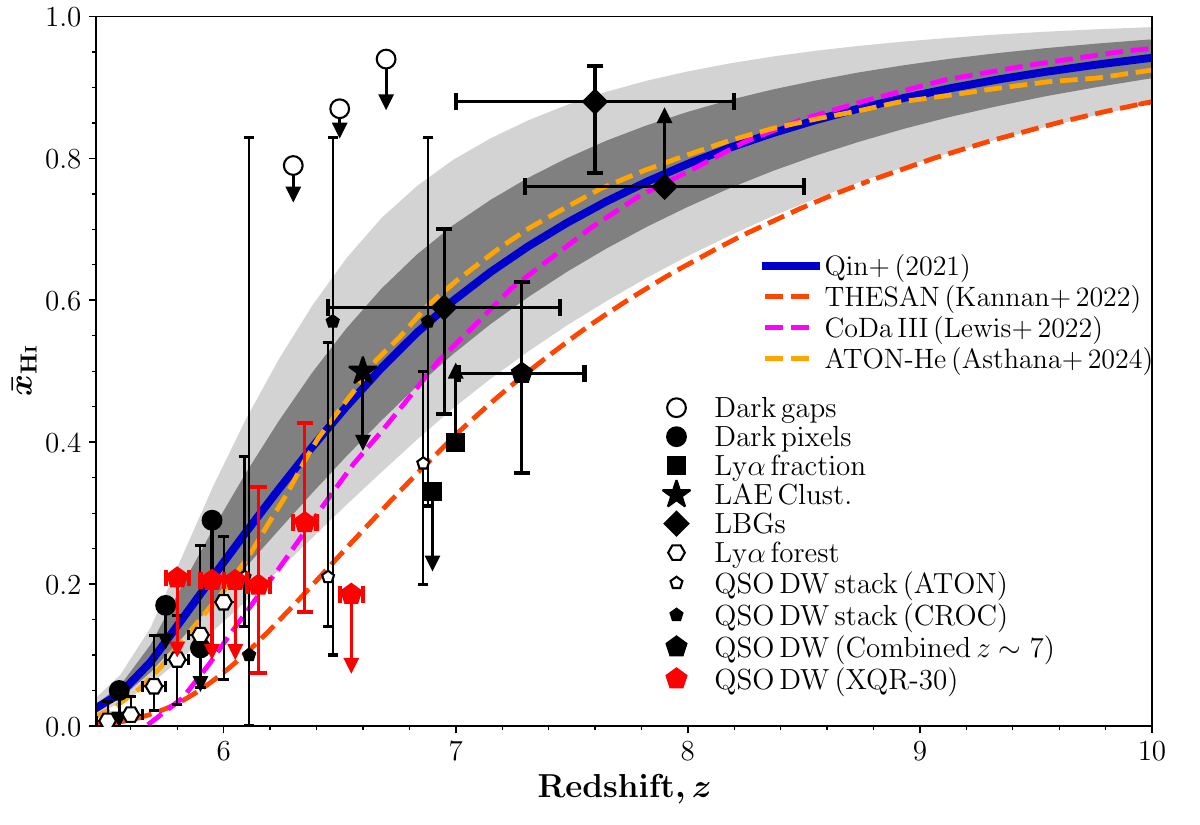}
	\end{center}
\caption[]{Our binned IGM neutral fraction constraints as a function of redshift in the context of other EoR observables. \textit{Red pentagons:} The XQR-30 sample (this work). \textit{Large black pentagon:} The combined IGM damping wing constraints of all 4 $z>7$ QSOs \citep{Greig:2022}. \textit{Small pentagons:} IGM damping wing constraints from stacking QSO reconstruction profiles \citep{Dominika:2024}. \textit{Open hexagon:} The inferred constraints from the \lya{} forest using numerical simulations at $z = 5.5-6.0$ \citep{Gaikwad:2023}. \textit{Open circles:} Dark gaps at $z=6.3, 6.5$ and 6.7 \citep[][obtained with XQR-30]{Zhu:2022}. \textit{Circles:} Dark pixels at $z=5.9$ \citep{McGreer:2015p3668} and $z=5.55, 5.75$ and 5.95 \citep{Jin:2023}. \textit{Squares:} the \lya{} fraction at $z=6.9$ \citep{Wold:2021} and $z=7$ \citep{Mesinger:2015p1584}, \textit{Stars:} LAE clustering at $z=6.6$ \citep{Sobacchi:2015}, \textit{Diamonds:} LBGs at $z=7$ \citep{Mason:2018}, $z=7.6$ \citep{Hoag:2019} and $z=8$ \citep{Mason:2019}. The blue curve and the dark and light shaded regions corresponds to the median, 1 and 2$\sigma$ constraints from \citep{Qin:2021} (see text for further details) and the dashed red, magenta and orange curves correspond to the fiducial reionisation histories of state-of-the-art numerical simulations of the IGM from THESAN \citep{Kannan:2022,Garaldi:2022}, CoDA III \citep{Lewis:2022} and \citet{Asthana:2024}.}
\label{fig:BinnedHistory}
\end{figure*}

To place our constraints in a broader reionisation context, in Figure~\ref{fig:BinnedHistory} we compare our binned IGM neutral fraction constraints against a compilation of other existing constraints on the IGM neutral fraction. Here, we consider constraints and limits obtained from: (i) dark gaps \citep{Zhu:2022} (ii) dark pixels \citep{McGreer:2015p3668,Jin:2023}, (iii) the \lya{} fraction at $z=6.9$ \citep{Wold:2021} and at $z=7$ \citep{Mesinger:2015p1584}, (iv) the clustering of \lya{} emitters (LAEs) at $z=6.6$ \citep{Sobacchi:2015}, (v) Lyman-break galaxies (LBGs) at $z=7$ \citep{Mason:2018}, $z=7.6$ \citep{Hoag:2019} and at $z=8$ \citep{Mason:2019}, (vi) the inferred constraints comparing numerical simulations to the \lya{} forest at $z = 5.5-6.0$ \citep{Gaikwad:2023}, (vii) the joint QSO IGM damping constraint combining all 4 $z>7$ QSOs \citep{Greig:2022} and (viii) IGM damping constraints after stacking continuum normalised QSO spectra \citep{Dominika:2024}. Note, this is not intended to be an exhaustive list, but instead provide a broad selection of approaches and redshift ranges. We also include the reionisation histories of the various state-of-the-art numerical simulations as introduced previously.

Our constraints on the IGM neutral fraction after binning in redshift intervals of $\Delta z = 0.1$ remain consistent with existing constraints on the EoR in the literature. Except for our limit at $z\sim6.5$, which is now inconsistent at more than 95 per cent. However, it is consistent with the recent QSO damping wing analysis of \citet{Dominika:2024}, although their posteriors are extremely broad owing to the parameter space binning of their simulations. The reason for our increased inconsistency is due to the multiplication of the individual posteriors within this redshift bin. Since three of the four QSOs are consistent with no IGM attenuation, the multiplication of these strengthens the preference for a posterior consistent with no IGM attenuation. In order to confirm the results of this work, independent re-analysis of the XQR-30 sample using alternative reconstruction and/or damping wing fitting are required.

\subsection{Discussion}

It is important to point out that the IGM damping wing constraints presented in this work use only a single QSO reconstruction methodology (covariance matrix). As highlighted previously, there are numerous approaches throughout the literature based on different underlying assumptions. For example, rather than explicitly using the correlations amongst emission lines to reconstruct the intrinsic flux, one can instead use correlations amongst the PCA components \citep[e.g.][]{Davies:2018a,Bosman:2021,Chen:2022}. Instead of directly mapping these correlations one can learn them via neural networks \citep{Dominika:2020} or deep learning \citep[][]{Liu:2021} and project the reconstruction uncertainties using normalising flows \citep{Reiman:2020}. Alternatively, one can use factor analysis to produce a more flexible basis set of components rather than those from PCA \citep{Sun:2022}. Finally, one can also perform deep learning to predict the QSO flux per pixel based on a training set of QSOs \citep{Fathivavsari:2020}.

In \citet{Greig:2024} we perform a detailed comparison of these reconstruction pipelines on a unified set of QSOs from two separate instruments (X-shooter and SDSS). Qualitatively speaking, these reconstruction pipelines tend to be consistent with one another within their associated modelling uncertainties. However, quantitatively, there can be some notable differences from object to object and these differences would filter through to attempts to recover constraints on the IGM neutral fraction through the damping wing. In the case of this dataset, with the high-quality X-shooter spectra, these numerical differences in the predicted QSO profiles could result in considerable differences in the inferred IGM neutral fractions from various reconstruction pipelines.

Not limited to just the reconstruction method, but the methodology to infer the IGM damping wing signature also differs in the literature. Here, we restrict our approach to fitting the smoothed IGM imprint using large-volume EoR simulations to generate synthetic damping wing profiles in the range $\lambda = 1218-1230$~\AA. Our choice is based on a preference to avoid having to model and marginalise over the uncertainties of the host QSO environment. In taking into account the QSO host environment, one can better deal with uncertainties in the QSO lifetime and affords a larger region over which to search for the damping wing imprint (e.g. $\lambda\sim1190-1230$\AA\ in \citealt{Davies:2018a}). Alternatively, one can simply use an analytic model of the red damping wing imprint \citep{MiraldaEscude:1998p1041} in order to gain an estimate of the IGM neutral fraction \citep[e.g.][]{Banados:2018,Reiman:2020}.

All of the approaches mentioned previously have their own underlying assumptions and uncertainties. Therefore, to be able to obtain robust estimates of the IGM neutral fraction from the smooth imprint of the IGM damping wing, realistically, one could average (marginalise) over the vast array of different approaches in the literature. In doing so, one would average over the differences in methodology and thus the various systematics and uncertainties. At the very least, independent confirmation using any of the existing other methods would add credence to our recovered constraints.

Another important point worth reiterating is that to be able to obtain our IGM damping wing constraints we had to (i) re-calibrate our reconstruction profile and (ii) degrade the observed X-shooter spectra. In the case of the former, this implies further improvements are necessary to our reconstruction pipeline to robustly account for the predicted \lya{} and \nv{} line profiles. In the case of the latter, we re-binned the spectra onto 1\AA\ bins and further had to increase the associated error in the spectra owing to the inflexibility of the analytic ($\chi^{2}$) expression used in our joint likelihood fitting. In addition to it being inflexible, it also does not fully take advantage of covariances in the observational data. Presently, the observed flux in each wavelength bin is take to be an independent measurement, ignoring the known correlations with neighbouring bins. In fact, we use these correlations to predict our reconstructed profiles (Gaussian components). Given the wealth of theoretical data at hand, one could employ a form of simulation based inference \citep[see e.g][for a recent review]{Cranmer:2020} to learn the likelihood that more directly connects our reconstructed QSO profiles to the underlying IGM neutral fraction of our synthetic damping wing profiles \citep[see e.g.][for a preliminary demonstration]{Chen:2023}.

Recently there have also been observations of \lya{} emission from galaxies extending into the EoR \citep[e.g.][]{Endsley:2022,Jung:2022,Hayes:2023,Saxena:2023,Umeda:2023,Whitler:2024,Witstok:2024}. Although galaxies are much fainter than QSOs, in principle a similar analysis can be performed to attempt to infer the presence of a \lya{} damping wing in the individual galaxy spectra \citep[see e.g.][]{Keating:2023}. However, this is extremely difficult as the local interstellar medium and \hi{} in the local circumgalactic medium make it much more challenging to infer the unabsorbed emission compared to that of QSOs. Improving the S/N or reducing the galaxy to galaxy variance through stacking \citep[e.g.][]{Umeda:2023} is fundamentally limited by the fact that the damping wing imprint is non-linear and cannot be described by a mean profile convolved with a mean galaxy spectral energy distribution (i.e. the average of a product is not the same as the product of averages).

\section{Conclusion} \label{sec:Conclusion}

We performed an IGM damping wing analysis on the enlarged XQR-30 sample, consisting of 42 high quality X-shooter spectra spanning $5.8 \lesssim z \lesssim 6.6$. Following careful selection cuts removing QSOs demonstrating broad absorption line features or possible proximate damped \lya{} absorption systems along the line of sight, we are left with 23 of the original 42 QSOs for our IGM damping wing analysis. Nevertheless, this is a factor of 5 improvement over the number of individual QSOs that have been explored for signs of IGM damping wing attenuation due to ongoing reionisation.

Our IGM damping wing analysis utilises a covariance matrix reconstruction approach to predict the intrinsic QSO profile near \lya{} \citep{Greig:2017}. The fundamental assumption of this approach is that emission lines can be accurately modelled as Gaussian profiles and that the \lya{} and \nv{} emission lines can be reconstructed from a covariance matrix of their correlations with other high ionisation emission lines. These lines, observed redward of \lya{} (e.g. \civ{}, \sioiv{} and \ciii{}), are easily measurable and unaffected by IGM attenuation or other line-of-sight contamination. In our approach we fit the observed QSO spectrum over $\lambda=1275-2300$\AA\ and draw reconstructed profiles spanning $\lambda=1180-1260$\AA\ from our resultant marginalised covariance matrix describing the properties of \lya{} and \nv{}.

In the process of analysing the XQR-30 sample, we identified a systematic offset in the predicted QSO flux between rest-frame \lya{} and \nv{} owing to our methodology of modelling emission lines as Gaussian profiles. Within this region, slight differences between the wavelength separation or in the widths of either the broad component of \lya{} or \nv{} can result in a very modest flux decrement. Using two distinct samples of QSOs from X-shooter and SDSS between $3.5 < z < 4.5$ (unaffected by IGM attenuation) we quantitatively explored whether this was specific to the higher quality X-shooter spectra, or a general feature of our pipeline. We found a consistent median flux decrement of $\sim10$~per cent across both samples, indicative of it being intrinsic to our reconstruction pipeline. Importantly, the consistency in the shape and amplitude of this flux decrement allowed us to recalibrate our covariance matrix reconstruction pipeline by drawing from templates of our original reconstructions compared to the known truth (unattenuated QSO spectrum). 

After recalibrating our reconstructed QSO profiles we jointly sample these with synthetic IGM damping wing profiles drawn from large volume EoR simulations (1.6 Gpc on a side with an EoR morphology driven by galaxies residing in $M_h \gsim 10^9 M_\odot$ haloes). Within our Bayesian framework we fit our reconstructed QSOs profiles multiplied by the synthetic IGM damping wing profiles against the observed QSO spectra of the XQR-30 sample. Specifically, we fit for the smooth component of the IGM damping wing imprint only redward of \lya{} ($\lambda=1218-1230$\AA). Following this pipeline, we recover 1D marginalised posteriors on the IGM neutral fraction from each individual QSO spectrum.

Across the available sample of 23 high-$z$ QSOs, we find 7 with constraints on the IGM neutral fraction during the EoR while for the remaining 16 we recover upper-limits. Making use of the large number of QSOs at our disposal, we binned our results in redshift intervals of $\Delta z = 0.1$. Following this, we obtain our final results on the IGM neutral fraction (median and 68th percentiles) during the tail-end of reionisation of $0.20\substack{+0.14 \\ -0.12}$, $0.29\substack{+0.14 \\ -0.13}$ at $z = 6.15$~and 6.35. Further, we report 68th percentile upper-limits on the IGM neutral fraction of $\bar{x}_{\hi{}}  < 0.21$, 0.20, 0.21 and 0.18 at $z = 5.8, 5.95, 6.05$~and 6.55.

These constraints on the IGM neutral fraction are consistent with those obtained from alternative methods in the literature. In particular, all but one of our constraints are within the joint 68th percentile constraints on the IGM neutral fraction obtained by the state-of-the-art forward-modelling of recent \lya{} forest data by \citet{Qin:2021}. The only discrepancy occurs in the highest redshift bin, $6.5 \leq z < 6.6$, which is inconsistent at 95 per cent certainty. Of the four QSOs from the XQR-30 sample in this redshift range, three are best modelled by upper-limits consistent with no IGM attenuation. The fourth (PSOJ323+12) indicates an IGM neutral fraction of $\bar{x}_{\hi{}} =0.23\substack{+0.19 \\ -0.15}$. Taken at face value this implies that these 4 QSOs may reside in a patch of the Universe that was reionised earlier. We also explored whether our choice of averaging over the adopted minimum local \hii{} bubble size, $R_{\rm min}$ containing the QSO plus galaxy contribution impacted these results. We found that increasing $R_{\rm min}$ had a fairly modest impact on the inferred value of $\bar{x}_{\hi{}}$, smaller than the corresponding 68th percentile uncertainties. Given the broad range of approaches in the literature designed to reconstruct the intrinsic properties of high-$z$ QSOs, along with differing methodologies for extracting the IGM damping wing imprint, it would be valuable to repeat this analysis across multiple methods, averaging over the different pipeline systematics and uncertainties.

\section*{Acknowledgements}

We would like to thank the anonymous referees whose comments improved this manuscript. Parts of this research were supported by the Australian Research Council Centre of Excellence for All Sky Astrophysics in 3 Dimensions (ASTRO 3D), through project number CE170100013. A.M. acknowledges support from the Ministry of Universities and Research (MUR) through the PRIN project ”Optimal inference from radio images of the epoch of reionization” as well as the PNRR project ”Centro Nazionale di Ricerca in High Performance Computing, Big Data e Quantum Computing”. G.D.B. was supported by the National Science Foundation (NSF) through grant AST-1751404. S.E.I.B. is supported by the Deutsche Forschungsgemeinschaft (DFG) under Emmy Noether grant number BO 5771/1-1. HC thanks the support by the Natural Sciences and Engineering Research Council of Canada (NSERC), funding reference \#DIS-2022-568580. MGH has been supported by STFC grants ST/N000927/1 and  ST/Y004191/1. Y.Z. was supported by the NSF through award SOSPADA-029 from the NRAO. This work was performed on the OzSTAR national facility at Swinburne University of Technology. The OzSTAR program receives funding in part from the Astronomy National Collaborative Research Infrastructure Strategy (NCRIS) allocation provided by the Australian Government, and from the Victorian Higher Education State Investment Fund (VHESIF) provided by the Victorian Government. Based on observations collected at the European Southern Observatory under ESO programmes: 60.A-9024, 084.A-0360, 084.A-0390, 085.A-0299, 086.A-0162, 086.A-0574, 088.A-0897, 091.C-0934, 294.A-5031, 096.A-0095, 096.A-0418, 097.B-1070, 098.B-0537, 0100.A-0625, 0101.B-0272, 0102.A-0154, 1103.A-0817.

Funding for the Sloan Digital Sky Survey IV has been provided by the Alfred P. Sloan Foundation, the U.S. Department of Energy Office of Science, and the Participating Institutions. SDSS acknowledges support and resources from the Center for High-Performance Computing at the University of Utah. The SDSS web site is \url{www.sdss4.org}.

SDSS is managed by the Astrophysical Research Consortium for the Participating Institutions of the SDSS Collaboration including the Brazilian Participation Group, the Carnegie Institution for Science, Carnegie Mellon University, Center for Astrophysics | Harvard \& Smithsonian (CfA), the Chilean Participation Group, the French Participation Group, Instituto de Astrofísica de Canarias, The Johns Hopkins University, Kavli Institute for the Physics and Mathematics of the Universe (IPMU) / University of Tokyo, the Korean Participation Group, Lawrence Berkeley National Laboratory, Leibniz Institut f{\"u}r Astrophysik Potsdam (AIP), Max-Planck-Institut f{\"u}r Astronomie (MPIA Heidelberg), Max-Planck-Institut f{\"u}r Astrophysik (MPA Garching), Max-Planck-Institut f{\"u}r Extraterrestrische Physik (MPE), National Astronomical Observatories of China, New Mexico State University, New York University, University of Notre Dame, Observat{\'o}rio Nacional / MCTI, The Ohio State University, Pennsylvania State University, Shanghai Astronomical Observatory, United Kingdom Participation Group, Universidad Nacional Aut{\'o}noma de M{\'e}xico, University of Arizona, University of Colorado Boulder, University of Oxford, University of Portsmouth, University of Utah, University of Virginia, University of Washington, University of Wisconsin, Vanderbilt University, and Yale University.

{\it Software:} \textsc{\small CYTHON} \citep{Behnel:2011}, \textsc{\small NUMPY} \citep{Harris:2020}, \textsc{\small SCIPY} \citep{Virtanen:2020}, \textsc{\small MATPLOTLIB} \citep{Hunter:2007} and \textsc{\small COSMOHAMMER} \citep{Akeret:2013}.

\section*{Data Availability}

The data underlying this article will be shared on reasonable request to the corresponding author.

\bibliography{Papers}

\appendix

\section{Updated Covariance Matrix} \label{sec:covariance}

\begin{figure*} 
	\begin{center}
	  \includegraphics[trim = 0.1cm 0.8cm 0cm 0.5cm, scale = 0.84]{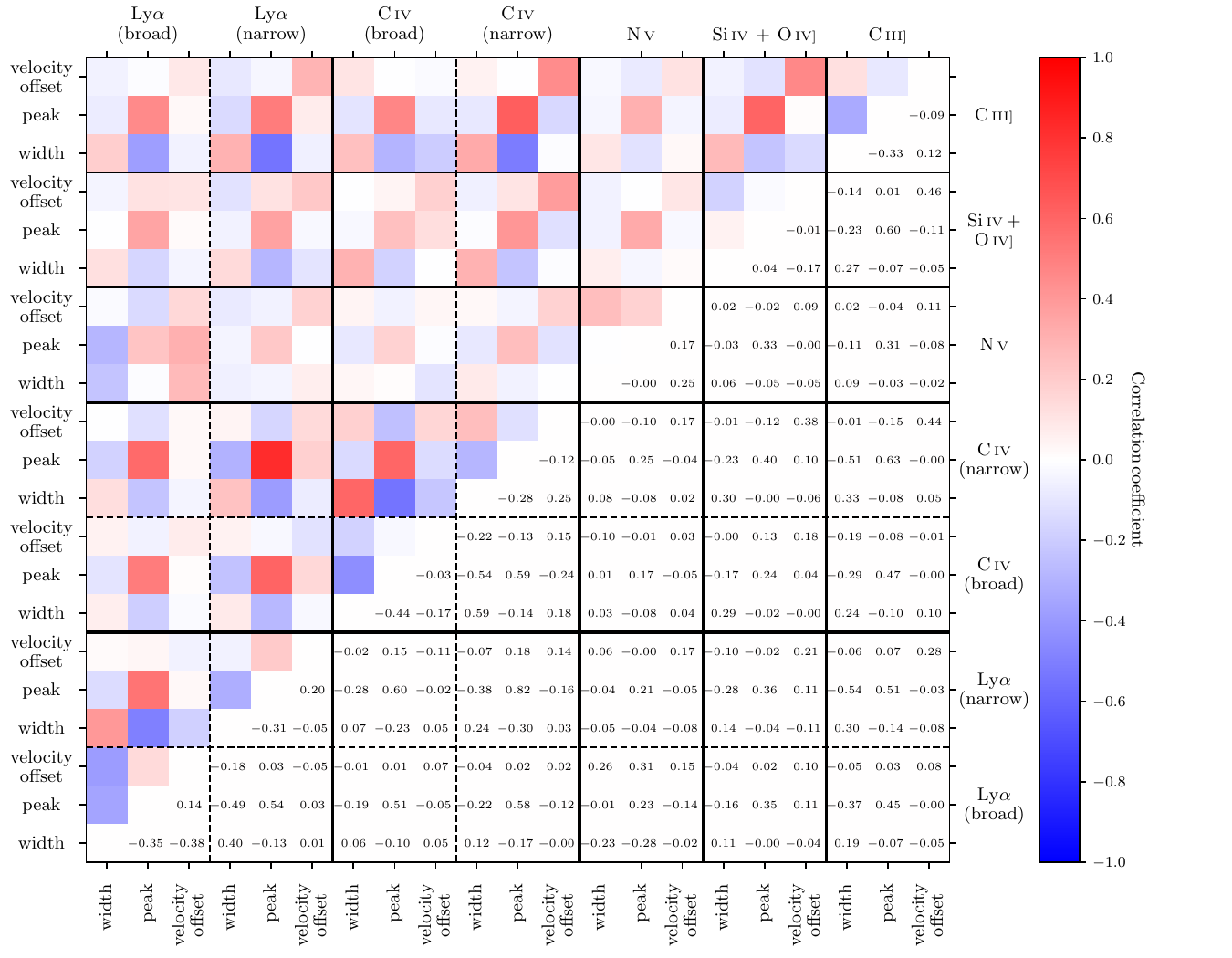}
	\end{center}
\caption[]{The updated correlation coefficient matrix following the expansion of the training set to 30,166 QSOs with S/N > 6.5 and spanning $2.08 < z < 4.0$ from BOSS DR16Q \citep{Lyke:2020}. This 21 dimensional matrix highlights the correlation coefficient strengths between the various emission line parameters (separated by solid black vertical and horizontal lines), with the \lya{} and \civ{} lines modelled by a double component Gaussian and the \nv{}, \sioiv{} and \ciii{} lines modelled as a single component Gaussian. Each emission line component is fully described by three parameters, the peak width, height and velocity offset from the systemic line centre. The dataset used in this work corresponds to an order of magnitude increase over previous work \citep{Greig:2022}.}
\label{fig:cov}
\end{figure*}

For this work we have substantially increased the number of QSOs that are included within our training set for constructing our emission line covariance matrix, from 1,673 QSOs to 30,166 QSOs. In Figure~\ref{fig:cov} we provide an updated correlation coefficient matrix for this new training set to demonstrate the available emission line parameter correlations. Each emission line is separated by solid vertical and horizontal dashed lines while dashed lines denote the separation between broad and narrow components, respectively. Along the top, each emission line is identified. 

This correlation coefficient matrix, $\textbfss{R}_{ij}$ is determined by computing, 
\begin{eqnarray}
\textbfss{R}_{ij} = \frac{\textbfss{C}_{ij}}{\sqrt{\textbfss{C}_{ii}\textbfss{C}_{jj}}},
\end{eqnarray}
where the $i$th and $j$th subscripts denote the different emission line parameters and $\textbfss{C}_{ij}$ is covariance matrix of the full training set computed using,
\begin{eqnarray}
\textbfss{C}_{ij} = \frac{1}{N-1}\sum^{N}_{i}(\textbfss{X}_{i} - \bmath{\mu}_{i})(\textbfss{X}_{j} - \bmath{\mu}_{j}).
\end{eqnarray}
Here, $\textbfss{X}_{i}$ is the data vector containing all $i$th emission line parameters (21) from the full QSO sample, $\bmath{\mu}$ is its mean and $N$ is the size of our training set. 

Generally speaking, we find only very small differences between the line correlation strengths from our updated training set compared to those shown in \citet{Greig:2022}.

\section{Impact of proximity zone size prior} \label{sec:prior}

Our constraints on the IGM neutral fraction are determined by averaging over four different minimum local \hii{} bubble sizes, $R_{\rm min}$ (see e.g. Section~\ref{sec:Damping} for further details) based on the available data used for our previously analysed $z>7$ QSOs. Here, we investigate this assumption of averaging over this $R_{\rm min}$ by determining the inferred IGM neutral fraction for a single QSO, PSOJ323+12 ($z=6.5872$) for each individual available $R_{\rm min}$ compared to that obtained when averaging over all four.

\begin{figure} 
	\begin{center}
	  \includegraphics[trim = 0.5cm 0.8cm 0cm 0.5cm, scale = 0.58]{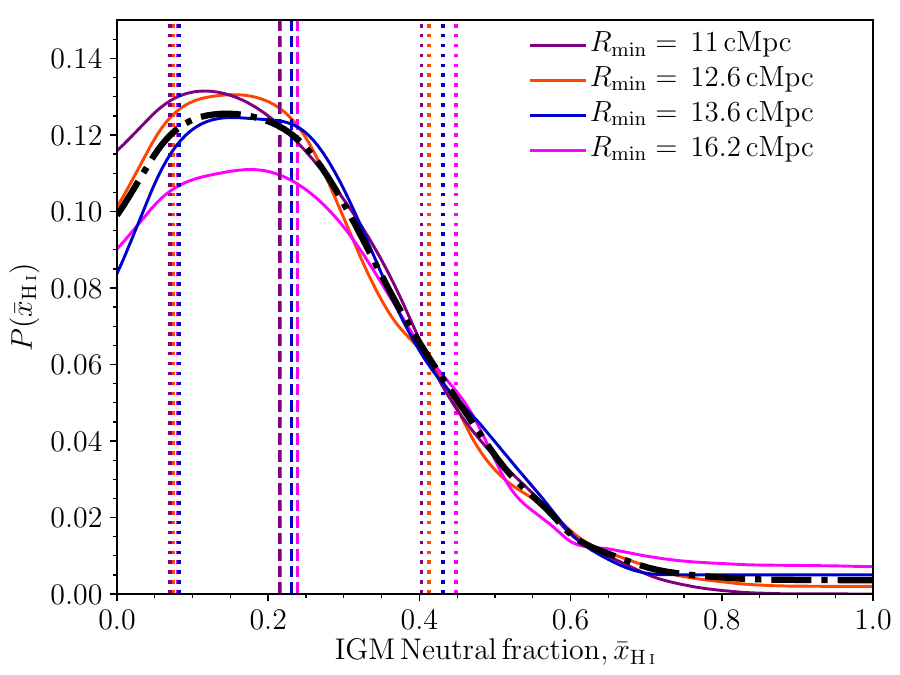}
	\end{center}
\caption[]{A comparison of the recovered IGM neutral fraction from PSOJ323+12 ($z=6.5872$) after averaging over different minimum \hii{} bubble sizes, $R_{\rm min}$ (black dot-dashed curve) and when considering each choice of $R_{\rm min}$ individually. In total we have data available for four choices of $R_{\rm min}$, distinguished by the coloured curves. The vertical dashed lines correspond to the median inferred constraint on the IGM neutral fraction for each individual choice of $R_{\rm min}$ while the dotted lines correspond to the 68th percentile region.}
\label{fig:pz}
\end{figure}

In Figure~\ref{fig:pz} we provide the inferred IGM neutral fraction PDF following our pipeline outlined in Section~\ref{sec:JointFitting} for an $R_{\rm min}$ of 11 (purple), 12.6 (red), 13.6 (blue) and 16.2 cMpc (magenta). The black dot dashed curve corresponds to averaging over all four choices while the vertical dashed lines correspond to the median of the recovered PDF whereas the dotted lines denote the 68th percentile uncertainties.

For these different available choices of $R_{\rm min}$ we obtain constraints of $\bar{x}_{\hi{}} = 0.21\substack{+0.18 \\ -0.14}$, $0.22\substack{+0.19 \\ -0.15}$, $0.23\substack{+0.20 \\ -0.15}$ and $0.24\substack{+0.21 \\ -0.16}$, respectively. For reference, after averaging we recover $\bar{x}_{\hi{}} =0.23\substack{+0.19 \\ -0.15}$. This implies that for an increasing choice of $R_{\rm min}$, we infer both an increasing IGM neutral fraction and also a broader PDF, which is to be expected. Increasing $R_{\rm min}$ effectively amounts to a marginal left-ward shift in the synthetic damping wing profiles in the left panel of Figure~\ref{fig:uncalibrated}. As the distance to the first patch of neutral IGM is larger (for the joint QSO + galaxy contribution), for the same fixed amplitude attenuation (determined by the reconstruction pipeline) we require higher IGM neutral fractions. Equally, due to the increased $R_{\rm min}$, we are sampling further into the tail of the \lya{} scattering cross-section, resulting in a lower amplitude variation for fixed wavelength in the synthetic damping wing profiles (reduced scatter in the mean profiles as a function of neutral fraction). This lower width between the damping wing profiles leads to an inferred broadening of the IGM posteriors as the width of our intrinsic profile distribution remains unchanged.

Overall, we recover a shift in the inferred IGM neutral fractions of $\Delta\bar{x}_{\hi{}} = 0.022$ for a shift in $R_{\rm min}$ of $\sim5$~cMpc. Additionally, we observe an increase of $0.02$ to the inferred 68th percentile region. Therefore, averaging over $R_{\rm min}$ does not have a strong impact on our inferred IGM neutral fraction constraints, especially given the overall width of the 68th percentiles in the posterior.

\bsp	
\label{lastpage}
\end{document}